\documentclass[hyper,a4paper]{JHEP}
\usepackage{amsmath,graphicx,epsfig,amssymb,multirow}
\allowdisplaybreaks

\preprint{LAPTH/1322\\ arXiv:0903.4705\\ \today}
\title{A model independent spin analysis of fundamental particles using
azimuthal asymmetries}
\author{Fawzi Boudjema \\
LAPTH, Univ. de Savoie, CNRS,\\
B.P. 110, F-74941, Annecy-Le-Vieux, FRANCE\\
Email: \email{boudjema@lapp.in2p3.fr}}
\author{Ritesh K. Singh \\
Institut f\"ur Theoretische Physik und Astrophysik, Universit\"at W\"urzburg,\\
D-97074  W\"urzburg, GERMANY\\
Email: \email{singh@physik.uni-wuerzburg.de}
}
\abstract{Exploiting the azimuthal angle dependence of the density
matrices we construct observables that directly measure the spin
of a heavy unstable particle.  A novelty of the approach is that
the analysis of the azimuthal angle dependence in a frame other
than the usual helicity frame offers  an independent cross-check
on the extraction of the spin. Moreover, in some instances when
the transverse polarisation tensor of highest rank is vanishing,
for an accidental or dynamical reason,  the standard azimuthal
asymmetries  vanish and would lead to a measurement with a wrong
spin assignment. In a frame such as the one we construct, the
correct spin assignment would however still be possible. The
method gives direct information about the spin of the particle
under consideration and the same event sample can be used to
identify the spins of each particle in a decay chain. A drawback
of the method is that it is instrumental only when the momenta of
the test particle can be reconstructed. However we hope that it
might still be of use in situations with only partial
reconstruction. We also derive the conditions on the production
and decay mechanisms for the spins, and hence the polarisations,
to be measured at a collider experiment. As an example for the use
of the method we consider the simultaneous reconstruction, at the
partonic level, of the spin of both the top and the $W$ in top
pair production in $e^+ e^-$ in the semi-leptonic channel.}
\begin{document}
\section{Introduction}
The Standard Model (SM) of particle physics has been successful in
explaining all the collider observables till date with a high
degree of precision. This is remarkable considering that the
particle content of the model is not complete since it requires a
scalar spin-less particle, the Higgs. In the SM formulation, this
particle is an integral ingredient of the mechanism of
electro-weak (EW) symmetry breaking. This mechanism is still not
well understood. For example, the fact that in the SM no symmetry
protects the mass of the spin-less Higgs poses the hierarchy
problem. Any solution to these issues brings in new particles and
interactions at TeV scale with varying spin and gauge quantum
number assignment. In a collider experiment, where almost all
these particles are expected to be produced and decay to the light
SM particles, the gauge quantum numbers, in principle, can be
re-constructed by adding-up the gauge quantum numbers of all the
observed light SM particles. Spin, on the other hand, shows up
only in the distributions in various kinematic variables in the
production and the decay sub-processes. Since the knowledge of the
spin, along with the gauge quantum numbers, can enable us to
distinguish amongst various candidate theories of physics beyond
the SM (BSM) there has been growing interest in this subject
recently in the context of the upcoming Large Hadron Collider
(LHC)~\cite{Choi:2002jk,
Barr:2004ze,Smillie:2005ar,Barr:2005dz,Meade:2006dw,Alves:2006df,Athanasiou:2006ef,Wang:2006hk,Smillie:2006cd,Kilic:2007zk,Alves:2007xt,Csaki:2007xm,Rajaraman:2007ae,Wang:2008sw,Kane:2008kw,Reinartz:2008zc,Osland:2008sy,Burns:2008cp,Alves:2008up}
and also in the context of proposed International Linear Collider
(ILC)~\cite{
AguilarSaavedra:2004ru,Battaglia:2005zf,Choi:2006mr,Buckley:2007th,Buckley:2008pp,Buckley:2008eb}.

Most of the BSM models have new particles that are partners of the
SM particles based on their gauge quantum numbers assignments.
However, the new particles may differ in the spin assignments. In
models with supersymmetry (SUSY), the spin of the SM partner
differs by $\frac{1}{2}$ owing to the fermionic nature of the SUSY
generators. There are however many other models such as UED where
the spin of the partner is same as in the SM. In both kind of
models a $Z_2$ symmetry can be left over, leading to a heavy
stable particle in the spectrum which can be the dark matter
candidate. In SUSY models, the lightest neutralino, singlino,
gravitino or axino can be stable, while in the models with
universal extra dimensions (UED) the first Kaluza-Klein (KK)
excitation of photon can be stable and are the dark matter
candidate. These dark matter candidates can not be detected
directly in collider experiments. Thus if these particles appear
at the end of a decay chain, the re-construction of spin can be
non-trivial, specially at a hadron collider like LHC. Moreover it
would be also important to infer the spin, and other properties,
of these dark matter candidates since these properties are
important for the indirect detection of dark matter in
astrophysical experiments.

The spin of a (new) particle determines the Lorentz structure of
its couplings with the other SM fermions and bosons. This, in a
way, fixes its dominant production and decay mechanisms. In many
cases a careful study of the energy dependence of the cross
section around threshold can distinguish between spin--$0$ and
spin--$\frac{1}{2}$ particles for example. Other methods to
determine spin involve decay particles correlators. At the heart
of these more direct methods is the decay helicity amplitude. For
example, the helicity amplitude of a particle with spin--$s$ and
helicity $\lambda$ with $-s \leq \lambda \leq s$ decaying into two
particles of spins $s_1$ and $s_2$, with helicity $l_{1,2}$
respectively, can be written as~\cite{Haber:1994pe}
\begin{equation}
M^{s\lambda}_{l_1l_2}(\theta,\phi) =\sqrt{\frac{2s+1}{4\pi}} D^{s*}_{\lambda l}
(\phi,\theta,-\phi) {\cal M}^s_{l_1,l_2} = \sqrt{\frac{2s+1}{4\pi}}
e^{i(\lambda-l)\phi}d^s_{\lambda l}(\theta) {\cal M}^s_{l_1,l_2}, \hspace{0.2cm} l=l_1-l_2.
\label{eq:M1to2}
\end{equation}
Here ${\cal M}^s_{l_1,l_2}$ is the reduced matrix element. This
has been written most conveniently in the rest frame of the
decaying particle. In fact the helicity here is the projection of
the spin on the quantisation axis. The polar angle $\theta$ is
measured w.r.t. this quantisation axis and the azimuthal angle $\phi$
is measured around the same quantisation axis with freedom to
chose the $\phi=0$ plane. In most of the examples, it is useful to
chose the production plane of the decaying particle as the
$\phi=0$ plane. Boosting along the quantisation axis will leave
the value of the helicity unchanged. The angular distribution in
these angles encodes the spin information through the rotation
matrix $D$ which {\em factorises} into an overall phase factor
$e^{i(\lambda-l)\phi}$ carrying the azimuthal angle $\phi$
dependence and the the $d^s_{\lambda l}$ function carrying the
polar angle dependence. The latter can be expressed
as~\cite{Brink-Satchler}
\begin{eqnarray}
  d^s_{\lambda l}(\theta )
  &=&  \sum_{t} (-1)^t \
        \frac{\left[(s+\lambda )! \ (s-\lambda )! \ (s+l)! \
                (s-l)!\right]^{1/2}}
        {t! \ (s+\lambda-t )! \ (s-l-t)! \ (t+l-\lambda)! }
    \nonumber\\
  &&    \times
        \left(\cos\frac{\theta}{2}\right)^{2(s-t)+\lambda-l}
        \left(\sin\frac{\theta}{2}\right)^{2t+l-\lambda}
\label{eq:ds}
\end{eqnarray}
with $-s \leq \lambda,l \leq s$. The sum is taken over all values
of $t$ which lead to non negative factorials.
The differential rates have therefore polynomial dependence on
$\cos\theta$ up to degree $2s$ and the azimuthal modulation coming
from the off-diagonal elements of the density matrix ranges up to
$\cos(2s\phi)$. One can construct observables to extract the
degree of $\cos\theta$ and/or $\cos\phi$ distribution. If the
highest mode for, say, the azimuthal dependence $\cos(2s\phi)$ can
be extracted this would be an unambiguous measure of the spin, $s$
of the particle.

Other methods have been used or advocated to determine   spin.
\begin{enumerate}
\item Exploiting the behaviour of the total cross-section
at threshold for pair production~\cite{Kane:2008kw,Choi:2006mr} or
the threshold behaviour in the off-shell decay of the
particle~\cite{Choi:2002jk},
\item distribution~\cite{Barr:2005dz,Athanasiou:2006ef,
Alves:2007xt,Rajaraman:2007ae,Choi:2006mr} in the production
(polar) angle relying on a known production mechanism,
\item comparing different spin assignments to intermediate particles in a
process for a given collider signature~\cite{Choi:2002jk,Meade:2006dw,Athanasiou:2006ef,Wang:2006hk,Smillie:2006cd,Kilic:2007zk,Osland:2008sy,Burns:2008cp,Alves:2008up},
\item comparing SUSY vs UED for a given collider signature~\cite{Smillie:2005ar,Barr:2005dz,Alves:2006df,Alves:2007xt,Csaki:2007xm,Wang:2008sw,Reinartz:2008zc,Battaglia:2005zf,Buckley:2007th,Buckley:2008pp,Buckley:2008eb},
\item extracting the $(\cos\theta)^{2s}$ polar angle dependence~\cite{Barr:2004ze,Smillie:2005ar,Athanasiou:2006ef,Wang:2006hk,Wang:2008sw,AguilarSaavedra:2004ru,Choi:2006mr,Lee:1958qu,Daumens:1975mn}
or $\cos 2 s\phi$ azimuthal angle
dependence~\cite{Buckley:2007th,Buckley:2008pp,Buckley:2008eb,Daumens:1975mn}
of the decay distributions.
\end{enumerate}
The first four methods are indirect ways to assess  the spin
information subject to some assumptions and can only support or
falsify a hypothesis. For example, the threshold behaviour depends
not only on the spin but the parity of the particle as
well~\cite{Choi:2002jk} and for a particle of given spin it could
be used to determine its parity~\cite{Bhupal Dev:2007is}. Further,
it has been shown~\cite{Choi:2006mr} that for pair production in
an $e^+e^-$ annihilation, the threshold  behaviour alone can not
determine the spin of the particle. With $\beta$ the velocity of
the produced particle in the laboratory,  at threshold the cross
section   for a scalar scales as  $\beta^3$ behaviour while for a
spin--$1/2$ it goes like $\beta$, except for Majorana fermions
which can have $\beta^3$ behaviour. Note that these $\beta^n$
characteristics do not take into account
Sommerfeld/Coulomb\cite{Sommerfeld} type corrections. The spin--1
particle can also have threshold behaviour and production angle
dependence same as that of fermions with only difference coming
from the distributions of their decay products. Thus, threshold
behaviour and production angle distributions can at best be used
only to confirm the spin assignment not to determine it. In the
second method, one usually assumes a production topology, like for
example $s$--channel pair production through a gauge boson.  In
this method the production angle dependence will depend upon the
spin of the test particle. But still this dependence is not unique
and can be obtained for higher spin test particles.

The third and the fourth method uses numerical values of
correlators or differences in the distributions, which can be
modified by the changes in the couplings or the presence of
addition particles in virtual exchange {\em etc}. Thus, one can
not use this method without having re-constructed the spectrum of
the theory experimentally. The last method, which uses decay
correlators, gives either the spin of the particle or the {\em
absolute lower limit} on its spin. We note that the moments of the
polar angle distribution discussed in Ref.\cite{Lee:1958qu} gives
an {\em upper limit} on the spin of the particle.

\subsection{Spin through the polar angle}
Earlier studies of spin measurements used the average values
of $\cos\theta$ or angular asymmetry, with appropriately defined
polar angle $\theta$, in the process of 2-body
decay~\cite{Lee:1958qu} or cascade decay~\cite{Daumens:1975mn}.
The numerical values of the angular asymmetries or the moments of
angular distribution gave estimates of the spin of the decaying
particles in a model independent way. Most of the recent spin
studies using decay kinematics focus on a decay chain that can be
realised in SUSY or UED models. All the intermediate particles in
the decay chains are assumed on-shell such that there is no
distortions coming from the shape of the off-shell propagator and
that it can be decomposed as a series of two body decays,
simplifying the calculations. For example, we look at a $3$--body
decay chain of a particle $A$, shown in Fig.\ref{fig:decaychain}.
We look in the rest frame of the intermediate particle $C$ whose
spin is to be determined. Using crossing symmetry we write the
matrix element for the $s$--channel process $AB\to C\to DE$
as~\cite{Haber:1994pe}
\begin{equation}
M^{\lambda_A,\lambda_B}_{\lambda_D,\lambda_E}(\theta_{BD},\phi)=(2s+1) \ \
d^s_{\lambda_i,\lambda_f}(\theta_{BD}) \ \ e^{i\phi(\lambda_i-\lambda_f)}
 \ \ {\cal M}^s_{\lambda_i,\lambda_f},
\end{equation}
where, $\lambda_i=\lambda_B-\lambda_A$ and $\lambda_f=\lambda_D-\lambda_E$.
The rotation matrix $d^s_{\lambda_i,\lambda_f}(\theta)$ for spin $s$ particle
is $2s$ degree polynomial in $\cos(\theta/2)$  and $\sin(\theta/2)$,
Eq.(\ref{eq:ds}), which on squaring transforms to a
$2s$ degree polynomial in $\cos\theta$. This leads to a $2s$
degree polynomial form of the angular distribution as
\begin{equation}
\frac{d\Gamma(A\to BDE)}{d\cos\theta_{BD}} = Q_0 + Q_1 \ \cos\theta_{BD}
+ \ ... \ + Q_{2s}  \  \cos^{2s}\theta_{BD}.
\label{eq:ctdist}
\end{equation}
Thus, we see that the degree of these polynomials is a consequence
of the representation of the particle under Lorentz or rotation
group, in other words, the spin of the particle, provided $C$ is
produced on-shell, i.e. $(p_D+p_E)^2 =p_C^2 = m_C^2 = $ constant.

\EPSFIGURE[ht]{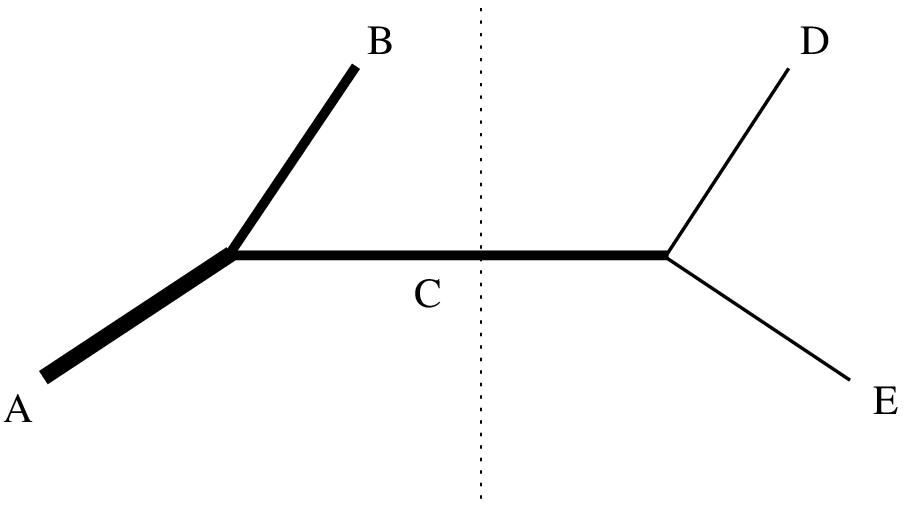}
{\label{fig:decaychain}The typical decay chain studied for the spin
analysis of particle $A$ via its decay into observed particles $B$ and $D$ and
a missing particle $E$.}

One can also describe the decay in powers of some invariants, the
highest power giving a measure of spin. Indeed, with
$m_{BD}^2=(p_B+p_D)^2$ we can write
\begin{equation}
\frac{d\Gamma(A\to BDE)}{dm^2_{BD}} = P_0 + P_1 \ m^2_{BD} + \ ... \ +
P_{2s}  \  (m^2_{BD})^{2s},
\label{eq:m2dist}
\end{equation}
obtained from Eq.(\ref{eq:ctdist}) through $dm^2_{BD} = 2E_B E_D
\beta_B \beta_D d\cos\theta_{BD}$ by using a transformation of
variables.  Note that we could have $P_{2s}=0$ and $P_{2s-1}\neq
0$ for some kinematical or dynamical reason, in this case we would
set the lower limit on the spin to be $s-\frac{1}{2}$. We note
that the above method involves two decay products of particle $A$
while it measures the spin of the intermediate particle $C$ and
not the spin of the mother particle $A$. To directly measure the
spin of $A$, we need to use the polar angle of every decay
products w.r.t. the quantisation axis of $A$. The distribution
w.r.t this decay angle looks identical to Eq.(\ref{eq:ctdist})
with $s$ being the spin of $A$.

\subsection{Spin through azimuthal angle}
Another method of direct spin re-construction is to use the
azimuthal angle distribution of the decay product about the
quantisation axis of the decaying particle $A$. This is the main
thrust of the present work. Using the form of the rotation
matrices it can be shown, see later, that the azimuthal
distribution appearing from the interference of different helicity
states, has the general form
\begin{equation}
\frac{d\Gamma}{d\phi} = a_0 + \sum_{j=1}^{2s} a_j \ \cos(j\phi)
+ \sum_{j=1}^{2s} b_j \ \sin(j\phi),
\label{eq:dphi_ab}
\end{equation}
with $a_j$ being the $CP$ even contributions while the $b_j$ being
$CP$ odd contributions.
A statistically significant non-zero value of $a_{2s}/a_0$ or
$b_{2s}/a_0$ proves the particle spin to be $s$. The coefficients,
$a_j$ and $b_j$, depend on the dynamics of production and decay
processes and we will see that they are proportional to the degree
of quantum interference of different helicity states of the
particle $A$, or in other words, to the off-diagonal elements of
production and decay density matrices. This distribution (but for
the $CP$ odd part) has been proposed in \cite{Buckley:2007th} and
used in \cite{Buckley:2008pp} to measure the spin of $W$ and $Z$
bosons at LEP-II and Tevatron, respectively. The azimuthal
distribution in the laboratory frame is not simple $\sin$ or
$\cos$, however it is sensitive to the polarisation of the
decaying particle as shown in Ref.~\cite{Godbole:2006tq}. In this
paper, we study the azimuthal distribution, Eq.(\ref{eq:dphi_ab}),
in a model independent way to determine the constants $a_j$s and
$b_j$s in terms of production and  decay mechanism and construct
collider observables to possibly measure these constants. We
construct the observables in two different frames of reference and
compare their merits.

This paper is organised as follows. In section~2 we give the
angular distribution of decay products for a general process of
production and decay with emphasis on the case of
spin--$\frac{1}{2}$ and spin--$1$ particles. We describe the
azimuthal distribution in terms of observables (asymmetries) to be
used at colliders or event-generators in section~3. A numerical
example of top quark decay chain is given in section~4 for the two
different reference frames. We conclude in section~5. Additional
expressions are given in the appendices.

\section{Density matrices, polarisation and azimuthal distributions}
To assess the spin of an unstable particle $A$, we look at a
general $n$--body production process $B_1 B_2 \to A \ A_1 \ ... \
A_{n-1}$ followed by the decay of $A$ as $A \to BC$, for example.
The other particles $A_i$'s produced in association with $A$ can
be either stable or decay inclusively. The differential rate for
such a process is given by (see for example
\cite{Godbole:2006tq}),
\begin{eqnarray}
d\sigma &=& \sum_{\lambda,\lambda'} \left[
\frac{(2\pi)^4}{2I} \rho(\lambda,\lambda')
\delta^4\bigg(k_{B_1}+k_{B_2}-p_A-\bigg(\sum_i^{n-1}p_i\bigg) \bigg)
\frac{d^3p_{A}}{2E_{A}(2\pi)^3} \ \prod_i^{n-1}
\frac{d^3p_{i}}{2E_{i}(2\pi)^3} \right]\nonumber \\
&\times&\left[ \ \frac{1}{\Gamma_A} \
\frac{(2\pi)^4}{2m_A} \Gamma'(\lambda,\lambda')\delta^4(p_A-p_B-p_C)
\frac{d^3p_{B}}{2E_{B}(2\pi)^3} \ \frac{d^3p_{C}}{2E_{C}(2\pi)^3}
 \right]
\label{eq:2tn}
\end{eqnarray}
after using the narrow-width approximation for the unstable
particle $A$, thereby factoring out the production part from the
decay. Here we have
$I^2=[m^2_{B_1B_2}-(m_{B_1}+m_{B_2})^2][m^2_{B_1B_2}-(m_{B_1}-
m_{B_2})^2]$, $m^2_{B_1B_2}=(k_{B_1}+k_{B_2})^2$,  $\Gamma_A$ is
the total decay width of $A$, $m_A$ is the mass of $A$ and
$\Gamma_A <<m_A$. The production and decay density matrices for
$A$ are denoted by $\rho(\lambda,\lambda')$ and
$\Gamma'(\lambda,\lambda')$, respectively. The terms in square
brackets in Eq.(\ref{eq:2tn}) are Lorentz invariant combinations.
The phase space integration can be performed in any frame of
reference without loss of generality.

Since we are interested in the decay distribution of $A$, we perform the
phase space integrations in the rest frame $A$. We integrate the first square
bracket in Eq.(\ref{eq:2tn}) and denote it as
\begin{equation}
\sigma(\lambda,\lambda') = \intop
\frac{(2\pi)^4}{2I} \rho(\lambda,\lambda')
\delta^4\bigg(k_{B_1}+k_{B_2}-p_A-\bigg(\sum_i^{n-1}p_i\bigg) \bigg)
\frac{d^3p_{A}}{2E_{A}(2\pi)^3} \ \prod_i^{n-1}
\frac{d^3p_{i}}{2E_{i}(2\pi)^3} \ .
\label{eq:prod}
\end{equation}
We note that the total integrated production cross-section,
without cuts, of the process is given by the sum of diagonal terms
$\sigma_{A} = {\rm Tr} \ \sigma(\lambda,\lambda')$, while the
off-diagonal terms of $\sigma(\lambda,\lambda')$ denote the
production rates for transverse/tensor polarisation states or, in other words,
for the quantum interference states. Further, we rewrite
$\sigma(\lambda,\lambda')= \sigma_A \ P_A(\lambda,\lambda')$,
where $P_A(\lambda,\lambda')$ is the polarisation density matrix
for $A$ in the corresponding production process. Similarly, we can
partially integrate the second term in Eq.(\ref{eq:2tn}) and write
it as
\begin{eqnarray}
\intop \frac{1}{\Gamma_A} \
\frac{(2\pi)^4}{2m_A} \Gamma'(\lambda,\lambda')\delta^4(p_A-p_B-p_C)
\frac{d^3p_{B}}{2E_{B}(2\pi)^3} \ \frac{d^3p_{C}}{2E_{C}(2\pi)^3} \nonumber\\
 = \frac{B_{BC} (2s+1)}{ 4\pi} \ \Gamma_A(\lambda,\lambda') d\Omega_B,
\label{eq:ddist}
\end{eqnarray}
where $B_{BC}$ is the branching ratio for the decay $A\to BC$, $s$
is spin of $A$, $\Gamma_A(\lambda,\lambda')$ is the decay density
matrix normalised to unit  trace, $d\Omega_B$ is the solid angle
measure for the decay product $B$.\footnote{One can also consider
3--body or higher body decay of $A$ in Eq.(\ref{eq:2tn}) and write
Eq.(\ref{eq:ddist}) by integrating all the phase space except
$\Omega_B$. One example of this will be the top-quark
decay~\cite{Godbole:2006tq}.} Combining Eq.(\ref{eq:prod}) and
Eq.(\ref{eq:ddist}) in  Eq.(\ref{eq:2tn}) we get the decay angular
distribution as
\begin{equation}
\frac{1}{\sigma} \
\frac{d\sigma}{d\Omega_B} = \frac{2s+1}{ 4\pi}
\sum_{\lambda,\lambda'} \ P_A(\lambda,\lambda') \ \ \Gamma_A(\lambda,\lambda'),
\label{eq:ndphi}
\end{equation}
where $\sigma = \sigma_A \ B_{BC}$, the total cross-section for
production of $A$ followed by its decay into $BC$ state. The
polarisation density matrix contains the dynamics of the
production process and we will discuss its form for
spin--$\frac{1}{2}$ and spin--$1$ particle in the following
sections. \\

\noindent First we will discuss the general structure of the decay
density matrix which can be studied independently of the
production mechanism.
The decay density matrix for a spin--$s$ particle, expressed in
terms of helicity amplitudes Eq.(\ref{eq:M1to2}), is given by
\begin{eqnarray}
\Gamma'^s(\lambda,\lambda')&=& \sum_{l_1,l_2} M^{s\lambda}_{l_1l_2}
M^{*s\lambda'}_{l_1l_2} \nonumber\\
&=& \left(\frac{2s+1}{4\pi}\right) e^{i(\lambda-\lambda')\phi}\sum_{l_1,l_2}
d^s_{\lambda l}(\theta) d^s_{\lambda'l}(\theta) \ |{\cal M}^s_{l_1,l_2}|^2 \nonumber \\
&=& e^{i(\lambda-\lambda')\phi} \sum_l d^s_{\lambda l}(\theta)
d^s_{\lambda'l}(\theta) \left[\sum_{l_1}
\left(\frac{2s+1}{4\pi}\right) |{\cal M}^s_{l_1,l_1-l}|^2  \right] \nonumber \\
&=& e^{i(\lambda-\lambda')\phi} \sum_l d^s_{\lambda l}(\theta)
d^s_{\lambda'l}(\theta) \ a^s_l, \label{eq:dgamma}
\end{eqnarray}
where
\begin{equation}
a^s_l=\left(\frac{2s+1}{4\pi}\right)\sum_{l_1} |{\cal M}^s_{l_1,l_1-l}|^2,
\hspace{1.0cm} |l_1| \le s_1, \ \ |l_1-l| \le s_2, \ \ |l| \le s.
\label{en:ajl}
\end{equation}
For a spin $s$ particle there are $2s+1$ different $a^s_l$'s that define the decay density matrix with
$$ {\rm Tr}(\Gamma'^s(\lambda,\lambda')) = \sum_{l} a^s_l \ .$$
Dividing the $\Gamma'^s$ by its trace leaves us with $2s$
independent quantities involving $a^s_l$'s to define the
normalised decay density matrix of a spin--$s$ particle. Now, the
normalised decay density matrix can be written as
\begin{equation}
\Gamma_A(\lambda,\lambda') = e^{i(\lambda-\lambda')\phi} \ \
\frac{\sum_l d^s_{\lambda l}(\theta)d^s_{\lambda'l}(\theta)
a^s_l}{\sum_{l} a^s_l} = e^{i(\lambda-\lambda')\phi} \
\gamma_A(\lambda,\lambda'; \theta), \label{eq:norm-ddm}
\end{equation}
where $\gamma_A(\lambda,\lambda'; \theta) \equiv
\gamma_A(\lambda,\lambda')$ is the {\em reduced} normalised decay
density matrix with only $\theta$ dependence left. It is important
to keep in mind that the $\phi$ dependence is an overall phase and
we see clearly that the differential cross section will have a
more transparent dependence on the azimuthal angle than the polar
angle. Using the relation in Eq.(\ref{eq:norm-ddm}) we can
re-write Eq.(\ref{eq:ndphi}) as
\begin{eqnarray}
\frac{1}{\sigma} \ \frac{d\sigma}{d\Omega_B}&=&\frac{2s+1}{ 4\pi}\left[
 \ \ \sum_{\lambda} \ P_A(\lambda,\lambda) \ \ \gamma_A(\lambda,\lambda)
\right.\nonumber\\
&+&\sum_{\lambda\neq\lambda'} \ \Re[P_A(\lambda,\lambda')] \ \ \gamma_A(
\lambda,\lambda') \ \cos((\lambda-\lambda')\phi) \nonumber\\
&-&\left.\sum_{\lambda\neq\lambda'} \ \Im[P_A(\lambda,\lambda')] \ \
\gamma_A( \lambda,\lambda') \ \sin((\lambda-\lambda')\phi)
\right],
\label{eq:dphi_rho}
\end{eqnarray}
which is similar to Eq.(\ref{eq:dphi_ab}) after integrating out
$\cos\theta$. Thus we have a simple looking $\phi$ distribution of
the decay product and also the coefficients of the different
harmonics in the distribution. We emphasise again that the $\phi$
dependence enters only through terms with $\lambda \neq\lambda'$,
in other words the off-diagonal elements of the production and
decay density matrices. When integrating over the full space
without any cuts, the information contained in these terms will be
lost. Another point to stress is that the form of the distribution
Eq.(\ref{eq:dphi_rho}) remains same in any other frame as long as
$\phi$ is measured around the momentum axis of the particle with
some suitable reference for $\phi=0$. The measurement of the $\cos
n \phi$ (with $n\leq 2s$) modulation that stems from the part
describing the decay  depends on the size of the corresponding
factor $P_A \gamma_A$ which is controlled by the interactions of
particle $A$. The factors $P_A$ describe the different
polarisations with which the particle is produced and the factor
$\gamma_A$ depends on the dynamics controlling the decay. One of
the aims of this paper is to analyse how one can use this
understanding to maximise these modulations and especially the
modulation with $\cos 2s \phi$ which is the most unambiguous
measure of the spin-$s$ of the decaying particle. For illustration
and  as a guide, in the following, we take a close look at the
production and decay density matrices for spin--$\frac{1}{2}$ and
spin--$1$ particles to identify the conditions on the production
and decay dynamics for the spin to be measured.

\subsection{Spin--${\frac{1}{2}}$ particle}
\label{sec:spinHalf} For the decay of spin--$\frac{1}{2}$
particle, $|\frac{1}{2},l\rangle \to |s_1,l_1\rangle +
|s_2,l_2\rangle$, the normalised decay density matrix, in the rest
frame or rather the {\em helicity rest
frame}~\cite{Leader:2001gr}, can be written as
\begin{eqnarray}
\renewcommand{\arraystretch}{1.5}
\Gamma_{\frac{1}{2}}(\lambda,\lambda') = \left[
\begin{tabular}{cc}
$\frac{1+\alpha\cos\theta}{2}$ & $\frac{\alpha\sin\theta}{2}
\ e^{i\phi}$\\
$\frac{\alpha\sin\theta}{2} \ e^{-i\phi}$ &$\frac{1-\alpha\cos\theta}{2}$
\end{tabular} \right],
\label{eq:gammahalf}
\end{eqnarray}
using Eq.(\ref{eq:dgamma}). Here $\alpha=(a^{1/2}_{1/2} -
a^{1/2}_{-1/2})/(a^{1/2}_{1/2}+a^{1/2}_{-1/2})$ and $a^j_l$ are defined in
terms of reduced matrix elements in Eq.(\ref{en:ajl}) and for
spin--$\frac{1}{2}$ particles given as
\begin{eqnarray}
a^{1/2}_{1/2}&=&\left(\frac{1}{2\pi}\right)\sum_{l_1}
|{\cal M}^{1/2}_{l_1,l_1-1/2}|^2  \hspace{1.0cm}
|l_1| \le s_1, \ \ |l_1-1/2| \le s_2 \nonumber\\
a^{1/2}_{-1/2}&=&\left(\frac{1}{2\pi}\right)\sum_{l_1}
|{\cal M}^{1/2}_{l_1,l_1+1/2}|^2  \hspace{1.0cm}
|l_1| \le s_1, \ \ |l_1+1/2| \le s_2.
\end{eqnarray}
The explicit calculation of $\alpha$ for the spin--$\frac{1}{2}$
particle decaying into two body final state is given in the
Appendices \ref{ap:ffv} and \ref{ap:ffs} for decays into a lighter
spin--$1/2$ and either a scalar or spin--$1$ considering general
effective operators. We have restricted ourselves to operators of
dimension 4. It can be seen that $\alpha$ is zero for a pure
vector or pure axial-vector coupling in the case of decay to a
spin-1. It can also be small depending on the masses of the
daughter particles.

The polarisation density matrix for a spin--$\frac{1}{2}$ particle
can be parameterised as
\begin{eqnarray}
\renewcommand{\arraystretch}{1.5}
P_{\frac{1}{2}}(\lambda,\lambda') = \frac{1}{2}\left[
\begin{tabular}{cc}
$1+\eta_3$ & $ \eta_1-i\eta_2$\\
$\eta_1+i\eta_2$ & $1-\eta_3$
\end{tabular} \right],
\label{eq:Phalf}
\end{eqnarray}
where $\eta_1$ is the transverse polarisation of $A$ in the
production plane, $\eta_2$ is the transverse polarisation of $A$
normal to the production plane and $\eta_3$ is the average
helicity or polarisation along the momentum of $A$ or polarisation
along the quantisation axis. Combining the expression of
$\Gamma_{\frac{1}{2}}(\lambda,\lambda')$ and
$P_{\frac{1}{2}}(\lambda, \lambda')$ in Eq.(\ref{eq:dphi_rho}) we
get the angular distribution of spin--$\frac{1}{2}$ particle
as~\cite{Godbole:2006tq}
\begin{equation}
\frac{1}{\sigma_1}\frac{d\sigma_1}{d\Omega_B} = \frac{1}{4\pi}
\left[ 1 + \alpha \eta_3 \cos\theta + \alpha \eta_1 \sin\theta
\cos\phi + \alpha \eta_2 \sin\theta\sin\phi  \right].
\label{eq:obsone}
\end{equation}
The $\cos\theta$ averaged azimuthal distribution is given by
\begin{equation}
\frac{1}{\sigma_1}\frac{d\sigma_1}{d\phi} = \frac{1}{2\pi} \left[ 1 +
\frac{\alpha\eta_1\pi}{4} \cos\phi
+ \frac{\alpha\eta_2\pi}{4} \sin\phi \right].
\label{fphi}
\end{equation}
Here we note that the $\cos\phi$ or the $\sin\phi$ modulation of
the azimuthal distribution is proportional to the transverse
polarisation of the spin--$\frac{1}{2}$ particles and also to the
{\em analysing power} $\alpha$. Thus, it is important that the
production process yields a non-zero value of either $\eta_1$ or
$\eta_2$. A non-zero $\eta_2$ indicates $CP$--violation or the
presence final state interaction (absorptive parts). A non-zero
$\eta_1$ can be obtained either with parity violation, which is
present in the electro-weak sector of the SM or with appropriate
initial beam polarisations. Further, we also need to know the
analysing power $\alpha$ of the particle. For spin--$\frac{1}{2}$
particle, it is given in Eqs.(\ref{eq:alffv}) and
(\ref{eq:alffs}). We see that the decay vertex has to be at least
partially chiral, {\it i.e.} parity violating for $\alpha\neq 0$.
That is, we need {\em effectively} chiral production and at least
{\em partially} chiral decay for the fermions for their spin to be
measured.

\noindent It is educative to realise that Eq.(\ref{eq:obsone}) can be cast
into
\begin{equation}
\frac{1}{\sigma_1}\frac{d\sigma_1}{d\Omega_B} = \frac{1}{4\pi}
\left[ 1 + \alpha \frac{\vec{p}_B}{|\vec{p}_B|}.\vec{\eta}
\right]. \quad {\rm with} \quad \vec{\eta}=(\eta_1,\eta_2,\eta_3)
\end{equation}
with $\vec{\eta}$ the polarisation vector. Performing a general
rotation will leave $\vec{p}_B.\vec{\eta}$ unchanged. In the new
frame, after rotation, we can define a new averaged azimuthal
distribution as
\begin{equation}
\frac{1}{\sigma_1}\frac{d\sigma_1}{d\phi^\prime} = \frac{1}{2\pi}
\left[ 1 + \frac{\alpha\eta_1^\prime\pi}{4} \cos\phi^\prime +
\frac{\alpha\eta_2^\prime\pi}{4} \sin\phi^\prime \right].
\end{equation}
If the rotation is done along the $\eta_2$ direction (normal to
the production plane), then $\eta_2^\prime=\eta_2$ but $\eta_1^\prime$
will pick up a contribution from $\eta_3$, the average helicity.
If $\eta_3 \gg \eta_1$ the azimuthal distribution in this new
frame is more conducive to a spin measurement, in the sense of
catching the $\cos \phi$ dependence.

It is important to observe that the picture we have described so
far in terms of azimuthal dependence through $\cos \phi$ and $\sin
\phi$ (or higher for higher spins) may be very much impacted if
cuts are applied to the cross section. If the cuts are
$\phi$-dependent, the azimuthal distributions may no longer have
the simple form of Eq.(\ref{fphi}) but would carry ``spurious"
dependence that would prevent the spin reconstruction as advocated
here. Indeed, we could have a much more complicated dependence of
the form
\begin{equation}
\frac{1}{\sigma_1}\frac{d\sigma_1}{d\phi} = \frac{1}{2\pi} \left[
F_c(\phi) + \frac{\alpha\eta_1\pi}{4} G_c(\phi) \cos\phi +
H_c(\phi)\frac{\alpha\eta_2\pi}{4} \sin\phi \right].
\end{equation}
unless only $\phi$ independent cuts are applied as suggested in
Ref.~\cite{Buckley:2008eb}.
\subsection{Spin--$1$ particle}
\label{sec:spin1} For the decay of spin--$1$ particle,
$|1,l\rangle \to |s_1,l_1\rangle + |s_2,l_2\rangle$, the
normalised decay density matrix is given by
\begin{equation}
\renewcommand{\arraystretch}{1.5}
\Gamma_1(l,l') = \left[
\begin{tabular}{ccc}
$\frac{1+\delta+(1-3\delta)\cos^2\theta+2\alpha \cos\theta}{4}$ &
$\frac{\sin\theta(\alpha+(1-3\delta)\cos\theta)}{2\sqrt{2}} \ e^{i\phi}$&
$(1-3\delta)\frac{(1-\cos^2\theta)}{4} \ e^{i2\phi}$\\
$\frac{\sin\theta(\alpha+(1-3\delta)\cos\theta)}{2\sqrt{2}} \ e^{-i\phi}$&
$\delta+(1-3\delta)\frac{\sin^2\theta}{2}$ &
$\frac{\sin\theta(\alpha-(1-3\delta)\cos\theta)}{2\sqrt{2}} \ e^{i\phi}$\\
$(1-3\delta)\frac{(1-\cos^2\theta)}{4} \ e^{-i2\phi}$ &
$\frac{\sin\theta(\alpha-(1-3\delta)\cos\theta)}{2\sqrt{2}} \ e^{-i\phi}$ &
$\frac{1+\delta+(1-3\delta)\cos^2\theta-2\alpha\cos\theta}{4}$
\end{tabular} \right],
\end{equation}
where,
\begin{equation}
\alpha  =  \frac{a^1_1 - a^1_{-1}}{a^1_1 + a^1_0 + a^1_{-1}} \hspace{1.0cm}
{\rm ,} \hspace{1.0cm}\delta=  \frac{a^1_0}{a^1_1 + a^1_0 + a^1_{-1}}
\end{equation}
and
\begin{eqnarray}
a^{1}_{1}&=&\left(\frac{3}{4\pi}\right)\sum_{l_1}
|{\cal M}^{1}_{l_1,l_1-1}|^2  \hspace{1.0cm}
|l_1| \le s_1, \ \ |l_1-1| \le s_2 \nonumber\\
a^{1}_{0}&=&\left(\frac{3}{4\pi}\right)\sum_{l_1}
|{\cal M}^{1}_{l_1,l_1}|^2  \hspace{1.0cm}
|l_1| \le \min(s_1,s_2) \nonumber\\
a^{1}_{-1}&=&\left(\frac{3}{4\pi}\right)\sum_{l_1}
|{\cal M}^{1}_{l_1,l_1+1}|^2  \hspace{1.0cm}
|l_1| \le s_1, \ \ |l_1+1| \le s_2 \nonumber\\
\end{eqnarray}
The explicit calculation of the analysing power parameter $\alpha$
(the vector part) and $\delta$ (the rank--$2$ tensor) for a
spin--$1$ particle decaying in two body final state is given in
the Appendices \ref{ap:vff}, \ref{ap:vvs} and \ref{ap:vvv}. In
particular $\delta=0$ for decays into massless fermions assuming
dimension--$4$operators. For the decay $W\to \bar ff'$ in the SM
we have $\alpha=-1$ for massless $f$ and $f'$.

The polarisation density matrix of spin--$1$ particle has two
parts: the vector polarisation which we define here as
$\vec{p}=(p_x,p_y,p_z)$ and is identical to that for a
spin--$\frac{1}{2}$ $\vec{\eta}$  and the tensor polarisation
described through a symmetric traceless rank--$2$ tensor $T_{ij}$,
${\rm Tr} \;T$=0. The density matrix is parameterised as~\cite{Leader:2001gr}
\begin{equation}
\renewcommand{\arraystretch}{1.5}
P_1(\lambda,\lambda') = \left[
\begin{tabular}{ccc}
$\frac{1}{3}+\frac{p_z}{2}+\frac{T_{zz}}{\sqrt{6}}$ &
$\frac{p_x -ip_y}{2\sqrt{2}}+\frac{T_{xz}-iT_{yz}}{\sqrt{3}}$ &
$\frac{T_{xx}-T_{yy}-2iT_{xy}}{\sqrt{6}}$ \\
$\frac{p_x +ip_y}{2\sqrt{2}}+\frac{T_{xz}+iT_{yz}}{\sqrt{3}}$ &
$\frac{1}{3}-\frac{2 T_{zz}}{\sqrt{6}}$ &
$\frac{p_x -ip_y}{2\sqrt{2}}-\frac{T_{xz}-iT_{yz}}{\sqrt{3}}$ \\
$\frac{T_{xx}-T_{yy}+2iT_{xy}}{\sqrt{6}}$ &
$\frac{p_x +ip_y}{2\sqrt{2}}-\frac{T_{xz}+iT_{yz}}{\sqrt{3}}$ &
$\frac{1}{3}-\frac{p_z}{2}+\frac{T_{zz}}{\sqrt{6}}$
\end{tabular}\right],
\end{equation}
Again using Eq.(\ref{eq:dphi_rho}) we can write the angular
distribution for spin--$1$ particle as
\begin{eqnarray}
\frac{1}{\sigma_2} \ \frac{d\sigma_2}{d\Omega_B} &=&\frac{3}{8\pi} \left[
\left(\frac{2}{3}-(1-3\delta) \ \frac{T_{zz}}{\sqrt{6}}\right) + \alpha \ p_z
\cos\theta+\sqrt{\frac{3}{2}}(1-3\delta) \ T_{zz} \cos^2\theta
\right. \nonumber\\
&+&\left(\alpha \ p_x + 2\sqrt{\frac{2}{3}} (1-3\delta)
 \ T_{xz} \cos\theta\right) \sin\theta \ \cos\phi \nonumber\\
&+&\left(\alpha \ p_y + 2\sqrt{\frac{2}{3}} (1-3\delta)
 \ T_{yz} \cos\theta\right) \sin\theta \ \sin\phi \nonumber\\
&+&(1-3\delta) \left(\frac{T_{xx}-T_{yy}}{\sqrt{6}} \right) \sin^2\theta
\cos(2\phi)\nonumber\\
&+&\left. \sqrt{\frac{2}{3}}(1-3\delta) \ T_{xy} \ \sin^2\theta \
\sin(2\phi) \right]. \label{disspinonept}
\end{eqnarray}
The $\cos\theta$ averaged distribution is
\begin{eqnarray}
\frac{1}{\sigma_2}\frac{d\sigma_2}{d\phi}&=&\frac{3}{4\pi} \left[
\frac{2}{3} + \frac{\alpha p_x \pi}{4} \cos\phi +
\frac{\alpha p_y \pi}{4} \sin\phi
+ \frac{2}{3}  (1-3\delta)\left(\frac{T_{xx}-T_{yy}}{\sqrt{6}}\right)\cos(2\phi)
\right.\nonumber\\
&&+ \left.\frac{2}{3} (1-3\delta)  \left( \sqrt{\frac{2}{3}} T_{xy}\right)
\sin(2\phi) \right].
\label{Vphi}
\end{eqnarray}
We note that the  $\phi$ and $2\phi$ modulation of the azimuthal
distribution is proportional to the transverse polarisations,
$p_x$ and $p_y$, and the transverse components of the tensor
polarisations, $T_{xx}-T_{yy}$ and $T_{xy}$. Again, in this frame
all the $\cos$ modulations are $CP$--even and the $\sin$
modulations are $CP$--odd. To determine the spin we need
$T_{xx}-T_{yy}\neq0$ from the production part and $\delta\neq1/3$
from the decay part in the $CP$--even production process. Since
there is no symmetry that sets $\delta$ to be one-third, the
dynamics in the decay part is not constrained. In other words, the
decay mechanism does not require any parity violation.

As we have done for spin--$1/2$ it is instructive to rewrite
Eq.(\ref{disspinonept}) in terms of invariants under rotations. If
one defines a rank--$2$ tensor out of the tensor product of the unit
vector describing the momentum of the decay product $B$,
$\mathbb{P}_B=\vec{p}_B\otimes
\vec{p}_B/|\vec{p}_B|^2\;\;$\footnote{$(\vec{p}_B\otimes
\vec{p}_B)_{ij}=p_{B\;i}\; p_{B\;j}$. The scalar product is
$T\mathbb{P}_B=\sum_{ij} T_{ij} \mathbb{P}_{B\;ij}={\rm Tr}\;T
\mathbb{P}_B $.}, using the fact that $T$ is traceless
Eq.(\ref{disspinonept}) writes in terms of (rotation) invariants
as
\begin{eqnarray}
\frac{1}{\sigma_2} \ \frac{d\sigma_2}{d\Omega_B} &=&
\frac{1}{4\pi} \left[ 1 + \alpha \ \frac{3}{2} \
\frac{\vec{p}_B}{|\vec{p}_B|}.\vec{p}+
(1-3\delta) \sqrt{\frac{3}{2}} \
{T}.\mathbb{P}_B \right]. \label{spinone-inv}
\end{eqnarray}
We can then rewrite Eq.(\ref{spinone-inv}) in another frame, in
particular one where we make a rotation around the $y$ axis,
transverse to the production plane. This will not mix the  CP-odd
and CP-even tensors but may make some $\phi$ asymmetries in the
new frame larger.

\subsection{spin--$\frac{3}{2}$ and spin--$2$}
For spin--$\frac{3}{2}$ and spin--$2$ particles we give the decay
density matrix in  Appendix~\ref{ap:dden}. Since we need the
coefficient of the highest harmonics to be non-zero for the spin
to be determined, we note that for spin--$\frac{3}{2}$ we need
parity violating interaction in the decay process, i.e., $\alpha_1
\neq 0$ and/or $\alpha_2\neq0$ whose combination defines the
analysing power of highest rank. For spin--$2$ particles we need
$A_4 \propto (a^2_2 - 4a^2_1 + 6a^2_0 -4a^2_{-1}+a^2_{-2}) \neq 0$.
$A_4$ is the analysing power of
rank--$4$, the highest rank for spin--$2$ to be non-vanishing (see
Eq.(\ref{eq:A4})). This can be achieved without parity violating
interactions in the decay process. We note that parity violating
interactions are required in the decay of fermions for its spin to
be measured along with its (transverse) polarisation being
non-zero . For the bosons, on the other hand, we only need its
transverse polarisation being non-zero either due to parity
violation in the production process or due to polarisation of the
initial beams.
%
%
\section{The azimuthal distribution at event-generators/colliders}
\label{sec:azimuthsx} The azimuthal distributions
Eqs.(\ref{fphi}), (\ref{Vphi}) {\it etc.} are given in the rest
frame of the decaying particle. To be able to measure the spin we
need to construct the above mentioned azimuthal angle in terms of
quantities defined in the lab frame of a collider experiment.
Before considering other frames let us first define some
asymmetries in the rest frame.

\subsection{Asymmetries in the rest frame}
 We start with
re-writing the rest frame azimuthal distribution in terms of some
simple asymmetries that we define below. Let us  first define
\begin{equation}
I_{2s}(\phi_1,\phi_2) = \intop_{\phi_1}^{\phi_2}
d\phi  \ \frac{d\sigma_{2s}}{d\phi}
\end{equation}
For $s=1/2$ we define the following asymmetries and calculate them
using Eq.(\ref{fphi}):
\begin{eqnarray}
A_1^1&=& \frac{I_1(-\pi/2,\pi/2) - I_1(\pi/2,3\pi/2)}{I_1(0,2\pi)}
= \frac{\alpha \eta_1}{2} \nonumber\\
B_1^1&=& \frac{I_1(0,\pi) - I_1(\pi,2\pi)}{I_1(0,2\pi)}
= \frac{\alpha \eta_2}{2}.
\label{A1I}
\end{eqnarray}
These asymmetries have been used in Ref.~\cite{Godbole:2006tq} as
a probe of the polarisation of the top-quark. Eq.(\ref{fphi}) can
be re-written in terms of these asymmetries as
\begin{equation}
\frac{1}{\sigma_1}\frac{d\sigma_1}{d\phi} = \frac{1}{2\pi} \left[ 1 +
\frac{\pi A_1^1}{2} \cos\phi
+ \frac{\pi B_1^1}{2} \sin\phi \right].
\label{fphiA}
\end{equation}
Similarly for $s=1$ we further define similar asymmetries and
calculate them in terms of vector and tensor polarisations as
follows,
\begin{eqnarray}
A_2^1&=& \frac{I_2(-\pi/2,\pi/2) - I_2(\pi/2,3\pi/2)}{I_1(0,2\pi)}
= \frac{3\alpha p_x}{4}\nonumber\\
B_2^1&=& \frac{I_2(0,\pi) - I_2(\pi,2\pi)}{I_2(0,2\pi)}
= \frac{3\alpha p_y}{4}\nonumber\\
A_2^2 &=& \frac{I_2(-\pi/4,\pi/4)-I_2(\pi/4,3\pi/4) +
I_2(3\pi/4,5\pi/4)-I_2(5\pi/4,7\pi/4)}{I_2(0,2\pi)}\nonumber\\
&&=\frac{2}{\pi} \ (1-3\delta) \ \left(\frac{T_{xx}-T_{yy}}{\sqrt{6}}
\right)\nonumber\\
B_2^2 &=& \frac{I_2(0,\pi/2)-I_2(\pi/2,\pi) +
I_2(\pi,3\pi/2)-I_2(3\pi/2,2\pi)}{I_2(0,2\pi)}\nonumber\\
&&=\frac{2}{\pi} \ (1-3\delta) \left(\sqrt{\frac{2}{3}}T_{xy}\right).
\label{A2I}
\end{eqnarray}
Thus  Eq.(\ref{Vphi}) can be re-written as
\begin{equation}
\frac{1}{\sigma_2}\frac{d\sigma_2}{d\phi} = \frac{1}{2\pi} \left[ 1 +
\frac{\pi A_2^1}{2} \cos\phi + \frac{\pi B_2^1}{2} \sin\phi +
\frac{\pi A_2^2}{2} \cos(2\phi) + \frac{\pi B_2^2}{2} \sin(2\phi)
\right].
\label{VphiA}
\end{equation}
We see that the $\phi$ distribution in the rest frame of the
decaying particle has very simple form in terms of the above
mentioned asymmetries. It is clear that for higher spins we need
to cut the $2\pi$ in more and more parts. The important
observation to make is that the coefficient of the $\cos \phi$,
$A_1^1$ for spin--$1/2$ and  $A_1^2$ for spin--$1$ are determined
exactly in the same way in terms of the asymmetries, this
generalises also in the same to higher spin particles and
similarly for other coefficients of $\cos j\phi$.

Next we write the asymmetries  in terms of spin-momentum correlators.
To this effect, we first  define the following spin vectors
in the helicity rest frame
\begin{equation}
s_x = (0,1,0,0), \ \ s_y = (0,0,1,0), \ \ s_z=(0,0,0,1)
\label{svec}
\end{equation}
which are orthogonal to the 4--momenta of the particle $A$,
$p_A=(m_A,0,0,0)$. These spin vectors satisfy the conditions
$p.s_i=0$ and $s_i.s_j=-\delta_{ij}$. The asymmetries for the
spin--$\frac{1}{2}$ case can then be written as
\begin{eqnarray}
A_1^1=\frac{\sigma(s_x.p_B <0)-\sigma(s_x.p_B >0)}{\sigma(s_x.p_B
<0)+\sigma(s_x.p_B >0)}, \ \ \ \ \
B_1^1=\frac{\sigma(s_y.p_B <0)-\sigma(s_y.p_B >0)}{\sigma(s_y.p_B
<0)+\sigma(s_y.p_B >0)}.
\label{A1sp}
\end{eqnarray}
The asymmetries for the spin--$1$ case can be written as
\begin{eqnarray}
A_2^1&=&\frac{\sigma(s_x.p_B <0)-\sigma(s_x.p_B >0)}{\sigma(s_x.p_B
<0)+\sigma(s_x.p_B >0)}, \ \ \ \ \
B_2^1=\frac{\sigma(s_y.p_B <0)-\sigma(s_y.p_B >0)}{\sigma(s_y.p_B
<0)+\sigma(s_y.p_B >0)},\nonumber \\[0.3cm]
A_2^2&=&\frac{\sigma( [s_x.p_B]^2-[s_y.p_B]^2 > 0) -
\sigma([s_x.p_B]^2-[s_y.p_B]^2 < 0)}{\sigma([s_x.p_B]^2-[s_y.p_B]^2 > 0) +
\sigma([s_x.p_B]^2-[s_y.p_B]^2 < 0)},\nonumber\\[0.3cm]
B_2^2 &=& \frac{\sigma( [s_x.p_B][s_y.p_B]>0)-\sigma( [s_x.p_B][s_y.p_B]<0)}
{\sigma( [s_x.p_B][s_y.p_B]>0)+\sigma( [s_x.p_B][s_y.p_B]<0)} \ \ .
\label{A2sp}
\end{eqnarray}
\TABLE[ht]{
\renewcommand{\arraystretch}{1.3}
\begin{tabular}{||c|c|l||}\hline\hline
Modulations & Asymmetries & Spin-momentum correlators ${\cal C}_j$\\\hline
$\cos(\phi)$  & $A^1$ & $s_x.p_B$ \\
$\sin(\phi)$  & $B^1$ & $s_y.p_B$ \\
$\cos(2\phi)$ & $A^2$ & $(s_x.p_B)^2-(s_y.p_B)^2 $ \\
$\sin(2\phi)$ & $B^2$ & $(s_x.p_B)(s_y.p_B) $ \\
$\cos(3\phi)$ & $A^3$ & $(s_x.p_B)^3-3(s_x.p_B)(s_y.p_B)^2 $ \\
$\sin(3\phi)$ & $B^3$ & $3(s_x.p_B)^2(s_y.p_B)-(s_y.p_B)^3$ \\
$\cos(4\phi)$ & $A^4$ & $(s_x.p_B)^4-6(s_x.p_B)^2(s_y.p_B)^2+(s_y.p_B)^4$ \\
$\sin(4\phi)$ & $B^4$ & $(s_x.p_B)^3(s_y.p_B)-(s_x.p_B)(s_y.p_B)^3$ \\
\hline\hline
\end{tabular}
\caption{\label{tab:cor}The table of asymmetries $A^j$ and $B^j$
corresponding to the $j\phi$ modulation of the azimuthal
distribution and corresponding spin-momentum correlators ${\cal C}_j$.
Here $s_{x,y}$ are the transverse spin directions of the decaying
particle $A$, with $s_x$ being in the production plane and $p_B$
is the 4--momentum of the decay product $B$. The spin vectors
$s_{x,y}$ are listed in Table \ref{tab:mom} in different frames.
For frame $F$ we replace $s_i$ with $\hat{s}_i$ in the above
correlators.
}}
It is clear from Eqs.(\ref{A1sp}) and (\ref{A2sp}) that the
asymmetry corresponding to a given modulation of $\phi$ has
identical expressions in terms of the spin-momentum correlator
$s_x.p_B$ and $s_y.p_B$. What we mean is that by constructing
specific functions with products of $s_i.p_B$ one reconstructs the
set of $\cos m\phi$, $\sin m \phi$, see Table~\ref{tab:cor}. Thus
in a spin independent way we can write these asymmetries as
\begin{equation}
A^j \ {\rm or} \ B^j = (-1)^j \ \frac{\sigma({\cal C}_j>0)-\sigma({\cal C}_j<0)}
{\sigma({\cal C}_j>0)+\sigma({\cal C}_j<0)} \ \ ,
\end{equation}
where the correlators ${\cal C}_j$s are listed in Table
\ref{tab:cor} for different modulations. Further we note that
these expressions of asymmetries have a simple interpretation in
terms of the polarisation parameters as long as $\vec{s_z}$,
3-vector, is parallel to the 3-momentum of the decaying particle
in the frame of choice and $\vec{s}_i$ are orthogonal to each
other. This defines a {\em helicity frame}. The lab frame,
achieved by a boost along $z$-axis and then rotation around
$y$-axis, also satisfies the properties of being a helicity frame.
We note that the orthogonality of $\vec{s}_i$ is respected only if
the boost is along one of  $\vec{s}_i$ directions. In some other
frame where these properties are not valid one needs to re-write
these $s_i$s as linear combinations of orthogonal  $s_i$ as we
will see in the following sections. This will be necessary
if the new frame is not reached through a boost made along
the direction of motion of the particle.
\EPSFIGURE[ht]{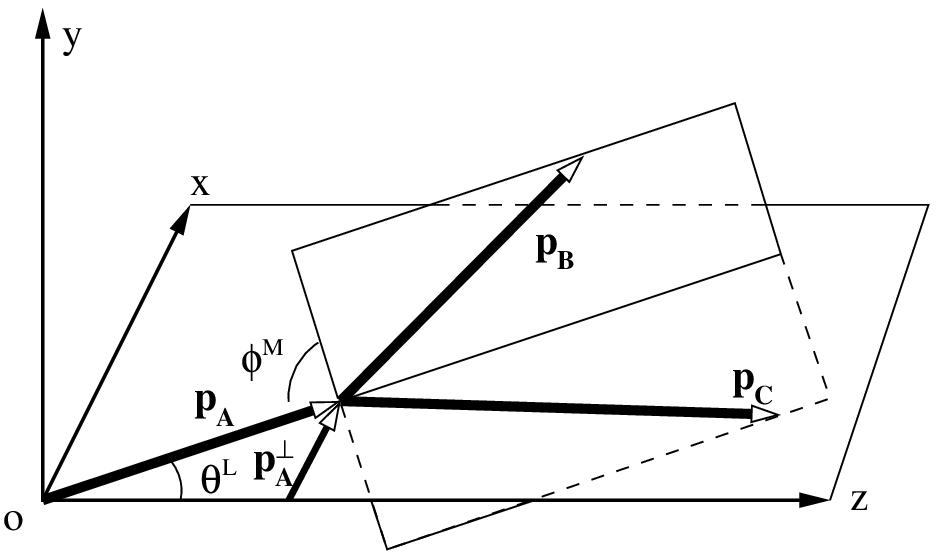}
{\label{fig:frameM}The momentum configuration in the {\em laboratory frame} $L$
of the colliding beams is depicted. The angle between the production
plane  (the $xz$--plane) and the decay plane (spanned by $p_B$ and $p_C$) is
the azimuthal angle $\phi^M$ ($=\phi^M_B$), which is the ordinary azimuthal
angle in the
frame $M$. This azimuthal angle has been studied in \cite{Buckley:2007th} and
\cite{Buckley:2008pp}.      }
\subsection{The rotated frame $M$}
The rotated frame is in fact the frame that is obtained from the
rest frame by performing a pure Lorentz boost along the
quantisation axis, the amount of boost is such that the energy of
the particle whose spin we want to study is the same as the one
measured in the laboratory frame. This is therefore a helicity
frame in the sense that the quantisation axis has now been
identified to lie along the momentum of the particle. The
normalised $3$--spin vectors $\vec{s_i}$ remain unchanged   and
therefore also the polarisation vectors ($\vec{\eta}$, $\vec{p}$)
and other tensor polarisations. Hence the azimuthal asymmetries
are the same as in the rest frame and have a  one-to-one
correspondance to the polarisation tensors defined in the rest
frame. The appellation rotated frame comes from how this frame is
picture in the laboratory frame. In fact this can be viewed as a
simple rotation. In the laboratory frame $L$, defined in Table
\ref{tab:mom}, the production plane of the particle $A$ defines
the $xz$--plane and the plane containing the decay products with
momenta $p_B$ and $p_C$ defines the decay plane. These two planes
intersect along the momentum $p_A$ of the decaying particle, see
Fig.~\ref{fig:frameM}. Thus the angle between these two planes is
the azimuthal angle of the decay product around the axis of spin
quantisation (the momentum $p_A$), {\it i.e.} the $\phi$ we have
mentioned in Eqs.(\ref{fphi}) and (\ref{Vphi}) and which
corresponds to exactly the azimuthal angle defined in the rest
frame. In terms of the variables defined in the laboratory frame
$L$, it is defined as~\cite{Buckley:2007th,Buckley:2008pp}
\begin{eqnarray}
\label{phim1} &&\phi = \cos^{-1} \left(\frac{(\hat{z}\times
\vec{p}_A^L).(\vec{p}_C^L\times \vec{p}_B^L)}
{|\hat{z}\times\vec{p}_A^L| \ |\vec{p}_C^L\times\vec{p}_B^L|}
\right) = \phi_B^M, \\
\label{phim2} {\rm where, \ \ }&& \phi_B^M = \tan^{-1} \
\frac{s_y^M.p_B^M}{s_x^M.p_B^M}=\tan^{-1} \
\frac{s_y^L.p_B^L}{s_x^L.p_B^L}.
\end{eqnarray}
From Eq.(\ref{phim1}) it  appears that the reconstruction of this
angle in the laboratory requires that one measures the momenta of
{\it all} the decay products. This may be achieved even if $C$,
say, is invisible provided  one has enough control and constraints
on the initial state so that the momentum of the decaying particle
$A$ is known, like for instance in $e^+ e^-$ annihilation where
the beam energy is fixed.

Another way to get the azimuthal angle $\phi_M$ it to re-construct
the scattering angle $\theta^L_A$ in the lab frame and rotate the
event about the $y$-axis by that angle to bring the momentum $p_A$
in the direction of the $z$-axis. The azimuthal angle of the decay
product, $\phi_B^M$ is same as $\phi$ mentioned above in the lab
frame. We dub this frame the {\em rotated frame} $M$ (obtained by
rotating the laboratory frame) and the momenta in this frame as
compared to that in the laboratory frame is given in the
Table~\ref{tab:mom}. Using the form of the momenta $p_B$ and $s_i$
in frames $R$(rest frame) and $M$, one can see that $\phi=\phi_B^R
=\phi_B^M$. Thus the distribution in angle $\phi$, defined in
Eq.(\ref{phim1}), is the same azimuthal distribution as in the
rest frame with same amplitudes for the different harmonics.

This azimuthal angle has been first studied in
Ref.~\cite{Buckley:2007th} to demonstrate the simple $\cos(j\phi)$
modulations of the azimuthal distributions and has been used to
examine the spin of $Z$ and $W$ bosons at Tevatron and LEP-II,
respectively, in Ref.~\cite{Buckley:2008pp}. Here we provide a
theoretical understanding of the amplitude of these $\cos(j\phi)$
modulations in terms of transverse polarisations of the particle
under consideration and its analysing power $\alpha$ {\it etc..}
In the event when the transverse polarisation is negligibly small
or zero, this frame will not give any modulation in the azimuthal
distribution as shown in section \ref{sec:app} for top pair
production in $e^+ e^-$. To address this potential issue we
construct another frame  which will give us an independent
estimate on the modulations of the $\phi$ distribution and hence
the spin of the particle. A hint on how to achieve this has been
illustrated in section~2.1 and section~2.2 for the spin--$1/2$ and
spin--$1$ when a simple rotation mixed the longitudinal
polarisation and the transverse polarisation in the production
plane, leaving the polarisation transverse to the production plane
unchanged. The next section will show how this can be achieved in
general and how we can construct the spin-momentum correlators in
this case.
\TABLE[ht]{
\renewcommand{\arraystretch}{1.3}
\begin{tabular}{||l|l||}\hline\hline
Rest frame := $R$ & Lab frame := $L$ \\ \hline
\begin{tabular}{l}
$s_x^R=(0,1,0,0)$\\ $s_y^R=(0,0,1,0)$ \\ $s_z^R=(0,0,0,1)$ \\
$p^R_A=(m_A,0,0,0)$ \end{tabular} &
\begin{tabular}{l}
$s_x^L=(0,\cos\theta^L_A,0,-\sin\theta^L_A)$\\ $s_y^L=(0,0,1,0)$ \\
$s_z^L=(\beta^L_A,\sin\theta^L_A,0,\cos\theta^L_A)\gamma^L_A$ \\
$p^L_A=E^L_A(1,\beta^L_A\sin\theta^L_A,0,\beta^L_A\cos\theta^L_A)$
\end{tabular} \\
$p_B^R=E_B^R \left(\begin{tabular}{c}
$1$\\ $\beta_B^R\sin\theta_B^R\cos\phi_B^R$ \\ $\beta_B^R\sin\theta_B^R
\sin\phi_B^R$\\ $\beta_B^R\cos\theta_B^R$
\end{tabular}\right)$ &
$p_B^L=E_B^L \left(\begin{tabular}{c} $1$\\
$\beta_B^L\sin\theta_B^L\cos\phi_B^L$ \\ $\beta_B^L\sin\theta_B^L
\sin\phi_B^L$\\ $\beta_B^L\cos\theta_B^L$
\end{tabular}\right)$ \\
\hline\hline
Rotated frame := $M$ & Boosted frame := $F$\\ \hline
\begin{tabular}{l}
$s_x^M=(0,1,0,0)$\\$s_y^M=(0,0,1,0)$\\$s_z^M=(\beta^L_A,0,0,1)\gamma^L_A$
\end{tabular} &
$\left.\begin{tabular}{l}
$\hat{s}_x^F=(0,1,0,0)$\\ $\hat{s}_y^F=(0,0,1,0)$ \\
$\hat{s}_z^F=(\beta_A^F,0,0,1)\gamma_A^F$
\end{tabular}\right\}$
\begin{minipage}{2.2cm}
{\scriptsize not the result of\\[-0.2cm] boost $L \rightarrow F$,\\[-0.2cm]
see text.}
\end{minipage}\\
$p^M_A=E_A^L(1,0,0,\beta^L_A)$ & $p^F_A=E^F_A(1,0,0,\beta_A^F)$ \\
$p_B^M=E_B^L \left(\begin{tabular}{c}
$1$\\ $\beta_B^L\sin\theta_B^M\cos\phi_B^M$ \\ $\beta_B^L\sin\theta_B^M
\sin\phi_B^M$\\ $\beta_B^L\cos\theta_B^M$
\end{tabular}\right)$  &
$p_B^F=E_B^F \left(\begin{tabular}{c}
$1$\\ $\beta_B^F\sin\theta_B^F\cos\phi_B^F$ \\ $\beta_B^F\sin\theta_B^F
\sin\phi_B^F$\\ $\beta_B^F\cos\theta_B^F$
\end{tabular}\right)$ \\
\hline\hline
\end{tabular}
\caption{\label{tab:mom}Momentum $p_A$, $p_B$ and the spin
directions $s_i$ in various frames. The transformation $R\to M$ is
a boost along $z$-axis $\Lambda_z(\beta^L_A)$, $M \to L$ is a
rotation $R(\theta^L_A)$ and $p^F_A$ \& $p_B^F$ are obtained by a
boost along $x$-axis $\Lambda_x(-\beta^L_A\sin\theta^L_A)$ from
frame $L$. Note that the spin vectors $\hat{s}^F_i$ in frame $F$
are not related to $s^L_i$ through boost but constructed such that
they represent the helicity basis. The expressions for
$s_i^F=\Lambda_x(-\beta^L_A\sin\theta^L_A)s_i^L$ which are the
result of the boost are given in Eq.~\ref{eq:hatteds}. The
azimuthal angle of  interest is $\phi =
\tan^{-1}(s_y.p_B/s_x.p_B)$ in each frame (with $s_i$ replaced by
$\hat{s}_i$ in frame $F$).} }

\subsection{The boosted frame $F$}
The idea behind the boosted frame $F$ is to induce a non zero
azimuthal asymmetry even in the event that transverse tensor
polarisations are very small or vanishing by making the
longitudinal components, assuming it is non zero, contribute. We
will show how this can be achieved especially how we can construct
the correlators from combinations of variables measured in the
laboratory frame. It should be added that  both the rest frame and
the $M$ frame are helicity frames. In the new frame and in order
to arrive at the mixing between the longitudinal and the
transverse polarisations we need to perform a transformation that
will move the longitudinal spin (quantisation axis) away from the
momentum of the particle. Yet, we still need to reconstruct a
helicity basis in order to construct the helicity density matrix.
To achieve the {\em misalignment}, we observe that a Lorentz boost
in a direction other than the direction of the particle momentum
will mix the helicity states. In our case one way to achieve this
is to carry a boost from the laboratory frame along the negative
$x$-axis with velocity $\beta^L_A\sin \theta^L_A$, thus reaching
the {\em boosted frame} $F$.  The momentum $p^F_A$ of $A$ is  then
pointing along the $z$-axis, see Table~\ref{tab:mom}. This looks
as if we have slowed down the particle, however contrary to frame
$M$ where the momentum is also pointing in the $z$ direction, we
can check that none of the transformed spin vectors
$s_i^F=\Lambda_x(-\beta^L_A\sin\theta^L_A) \ s_i^L$ has its three
momentum lying on the momentum of the particle. In fact, it can be
shown on general ground that if initially the spin axis is
parallel to the particle momentum, in the new frame these two axes
will move away by an angle $\omega$, the Wick
angle~\cite{Leader:2001gr}, if the boost is not performed along
the particle momentum.
\EPSFIGURE[ht]{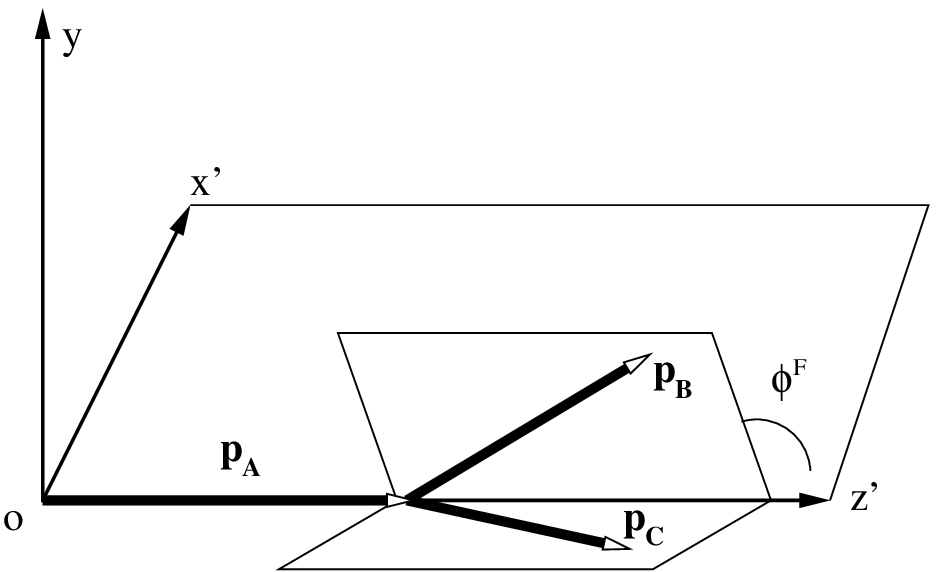}
{\label{fig:frameF}The momentum configuration in the transversely {\em boosted
frame} $F$ is depicted. The azimuthal angle of $p_B$ is  the azimuthal angle
$\phi^F$ ($=\phi_B^F$). This frame is obtained from frame $L$ by boosting
along negative $x$-axis such that the transverse momentum $p_A^\perp$ becomes
zero. }
Indeed we verify that
\begin{eqnarray}
s_x^F&=&\Lambda_x(-\beta^L_A\sin\theta^L_A) \; s_x^L=\cos \omega
\; \hat{s}_x^F
-\sin \omega \; \hat{s}_z^F\nonumber \\
s_y^F&=&\hat{s}_y^F=s_y^L \nonumber \\
s_z^F &=& \Lambda_x(-\beta^L_A \sin \theta^L_A)\ s_z^L=
\sin \omega \; \hat{s}_x^F  + \cos \omega \;\hat{s}_z^F
 \quad {\rm with}  \quad \nonumber \\
\cos \omega&=&\frac{\cos\theta_A^L}
{\sqrt{1-(\beta_A^L\sin\theta_A^L)^2}},  \quad \sin \omega=
\frac{\sin\theta_A^L}{\gamma_A^L \sqrt{1-
(\beta_A^L\sin\theta_A^L)^2}} \label{eq:hatteds}
\end{eqnarray}
$\hat{s}_{x,y,z}^F$ are the helicity basis in the new frame $F$
and are given explicitly in Table~\ref{tab:mom}. Since the spin
vectors $s_{x,z}^L$ in the laboratory frame are not parallel to
the $x$-axis (except for $\sin\theta^L_A=0$ or $1$), a boost along
the $x$-axis modifies the orthogonality of the spatial component
of $s_i$ in the boosted frame. Also the spatial components of the
boosted $s_z$, i.e. $\Lambda(-\beta^L_A\sin \theta^L_A) s_z^L$, is
not parallel to the spatial component of $p^F_A$, owing to the
Wick rotation of the spin basis. Thus the definition of the
longitudinal or transverse polarisations in the frame $F$, which
is not a helicity frame, is different from that in the helicity
lab frame $L$. Since the asymmetries $A^j$ and $B^j$ in the frame
$F$ are defined with respect to $\hat{s}_i^F$, we can in principle
have $A^j$ non-zero even in the absence of any transverse
polarisation in frame $L$. The helicity basis $\hat{s}_{x,y,z}^F$
is to be used to construct the spin-momentum correlators in frame
$F$.

\noindent This rotation of the spin basis vectors leads to the
transformation of the density matrix and the various polarisation
parameters. In general, for a rotation defined through the, Euler,
angle $\tilde{\theta}, \tilde{\phi}$  the density matrix
transforms as~\cite{Leader:2001gr}
$$\rho^\prime = D(\tilde{\phi},\tilde{\theta},-\tilde{\phi}) \
\rho^M \
D^\dagger(\tilde{\phi},\tilde{\theta},-\tilde{\phi}) \ .$$
In our case we have $\tilde{\theta}=\omega$ and
$\tilde{\phi}=0$(boost in the $x$ direction), which leads to
\begin{equation}
\rho^F(\lambda,\lambda') = d^s_{\lambda l}(\omega) \ d^s_{\lambda'
l'}(\theta_{\omega}) \ \rho^L(l,l') \ .
\end{equation}
Thus the density matrix $\rho^F$ does not receive any additional
phase and the azimuthal $\sin(n\phi_B^F)$ dependence remains unaltered.
The polarisation parameters, for the spin--$1$ for example as
concerns the vector and the tensor polarization, transform
as~\cite{Leader:2001gr}
\begin{equation}
p_i^F = R_{ij}(\omega) \ p_j^L \ , \hspace{0.5cm} T_{i,j}^F =
R_{ik}(\omega) \ R_{jk}(\omega) \ T_{kl}^L \label{eq:pT_F}
\end{equation}
as mentioned earlier in Sec.~\ref{sec:spinHalf} and
\ref{sec:spin1}. Here, $i,j,k,l = \{x,y,z\}$, $R_{ij}$ is the
 matrix, which for the boost we have performed, corresponds
to a rotation about the $y$-axis in the cartesian coordinate. The
superscript $F$ or $L$ stands for the quantities defined in frame
$F$ or frame $L$ respectively. The asymmetries $A^j$ and $B^j$ in
frame $F$ have exactly the same expression as Eqs(\ref{A1I}) and
(\ref{A2I}) with $p_i$ and $T_{ij}$ replaced by the ones defined
in Eq.(\ref{eq:pT_F}). Thus for a spin-$1$ particle we have $A^1
\propto p_x^F = R_{xj}(\omega) p_j^L$ and not simply related to
the transverse polarisation $p_x^L$ as we have in the (helicity)
frames $M$ or $L$. This shows again that one can have a non-zero
$A^1$ even in the absence of transverse polarisation in the frames
$M$ or $L$.

\noindent The azimuthal angle in this boosted frame is denoted by
$\phi_B^F$ and shown in Fig.\ref{fig:frameF}. Since $\phi_B^F$ is
the azimuthal angle around the new momentum $p^F_A$, it will have
a simple $\cos(j\phi)$ and $\sin(j\phi)$ modulations in the
distribution up to $j=2s$. The asymmetries $A^j$ and $B^j$ in this
frame are defined w.r.t. the spin directions $\hat{s}_i^F$ given
in the Table~\ref{tab:mom}. $\phi_B^F$ is expressed as
\begin{eqnarray}
\cot \phi_B^F &=& \frac{\hat{s}_x^F.p_B^F}{\hat{s}_y^F.p_B^F} =
\frac{\cos\omega \ s_x^L.p_B^L + \sin\omega \ s_z^L.p_B^L}
{s_y^L.p_B^L} \label{phif1}\\
&=& \cos\omega \ \cot\phi_B^M + \sin\omega  \
\frac{s_z^L.p_B^L}{s_y^L.p_B^L}. \label{phif2}
\end{eqnarray}
We see that $\phi_B^F$ is related to $\phi_B^M$ in a non-trivial
way and thus the corresponding modulation need not have vanishing
amplitudes even when this is the case in terms of $\phi_B^M$
distribution. Note that the asymmetries can be zero in both
frames if either we have $P_A(\lambda,\lambda')\propto
\delta_{\lambda, \lambda'}$, {\it i.e.} when the particle is
completely unpolarised, or when the particle is spin--$0$. In all
other cases, the two frames will lead to different values of the
azimuthal asymmetries. Thus, we need to use both frames to confirm
the spin of the particle.
%

\subsection{Note on event reconstruction}
The asymmetries $A^j$ and $B^j$ and the azimuthal angles $\phi^M$
and $\phi^F$ require complete reconstruction of the test
particle's momentum in order to construct the corresponding spin
vectors $s_i^L$ and/or $\hat{s}_i^{L/F}$. The possibility of
reconstruction depend both on the kind of collider and the number
of missing particles in the process. For example, reconstruction
of spin vectors is possible at colliders with {\em fixed} center
of mass energy, like the ILC, for some selected processes where
the number of missing particles is 2 or less. At hadronic
colliders, having variable centre of mass energy at partonic
level, such reconstructions can be achieved for processes with one
or no missing particles. Most of the new physics models with a
dark matter candidate have two missing particles in the production
process of new particles at LHC. This makes the desired
re-construction as outlined here unfortunately impossible at LHC
in such processes. It is worth further investigating how this
method could be combined with other methods  or improved. To
illustrate the method we therefore turn to an application for a
collider such as the ILC.

\section{Application to $e^+e^- \to t\bar t \to bW^+ \ \bar bW^- \to bl^+\nu
 \ \bar b j j$}
\label{sec:app} In this section we study  top-quark pair
production in $e^+e^-$ in the semi-leptonic channel as a test bed
for the spin measurement based on the exploitation of the
azimuthal asymmetries in different frames outlined previously. We
chose this particular process because it represents a decay chain
where the intermediate $W$ boson is on-shell. The charged lepton,
$l^+=e^+, \mu^+$, in the leptonic decay of the top (and the $W^+$)
will play the role of our particle $B$ in the previous section and
used to construct the spin-momentum correlators.
In this example all the momenta can be reconstructed and therefore
the methods we have outlined can be applied readily. We do not
take beamsstrahlung into account nor do we consider the issue of
backgrounds that might force us to introduce cuts, which we want
to avoid. However, we consider the effect of beam polarisation.
The polarisations of the initial electron and position beams can
be used to tune the polarisation of the produced heavy particles
that can drastically affect the polarisation. We work at
$\sqrt{s}=500$GeV where the total cross section, including
branching ratios, is $81$fb for unpolarised $e^+,e^-$, This
corresponds to a total number of $40500$ events with a typical
luminosity of $500$ fb$^{-1}$. For each fit we make, we will indicate
what the minimum number of events is required for a $3\sigma$
discovery of a particular $\cos j \phi$ modulation that is a
measure of the spin of the particle, we will see that this
programme could be successfully carried at a linear collider with
$500$ fb$^{-1}$ for this process.

For event generation we use the partonic level event generator
{\tt Pandora-2.3}~\cite{p23} and generate $2 \; 10^6$ events for
 different initial state polarisations.
Event by event we need to calculate $s_x^M.p_B^M,s_y^M.p_B^M$
(frame $M$) and $\hat{s}_x^F.p_B^F,\hat{s}_y^F.p_B^F$ (frame $F$)
which can be expressed in terms of energies and angles measured in
the lab frame,
\begin{eqnarray}
s_x^M.p_B^M &=& s_x^L.p_B^L=-E_B^L(\cos\theta_A^L
\sin\theta_B^L\cos\phi_B^L
- \sin\theta_A^L \cos\theta_B^L)\nonumber \\
\hat{s}_x^F.p_B^F&=&\frac{E_B^L(\beta_A^L\sin\theta_A^L-\sin\theta_B^L
\cos\phi_B^L)}
{\sqrt{1-(\beta_A^L\sin\theta_A^L)^2}} \nonumber\\
s_y^M.p_B^M &=&= \hat{s}_y^F.p_B^F=s_y^L.p_B^L=
-E_B^L\sin\theta_B^L \sin\phi_B^L \ .\label{sxpand}
\end{eqnarray}
 We then calculate, for $A=t, W$ all the 8
asymmetries corresponding to the correlators in Table~\ref{tab:cor},
therefore testing whether a value for the top spin as high as $s=2$ is
possible. The azimuthal angles in
frames $M$ and $F$ can be constructed using Eq.(\ref{sxpand}) along with
Eqs.(\ref{phim2}) and (\ref{phif1}) for generating the distributions.
The reconstructed azimuthal distributions are then fitted
with a general function
\begin{equation}
F_n(\phi) = a_0 + \sum_{j=1}^{n} \left[ a_j \cos(j\phi) + b_j
\sin(j\phi) \right] \label{eq:fita}
\end{equation}
with $n=4$. With $n=4$, the only bias is that the particle has
spin $s\leq 2$. We then compare the best fit coefficients with the
asymmetries calculated. Since we work with the SM production and
decay mechanisms for the $t$--quark, there is no $CP$ violation in
this process. The fitting procedure returns $b_j\approx 0$ and
$B^j\approx 0$ in both  frames $M$ and $F$ for all the initial
state polarisations.  This constitutes therefore a consistency
check and confirms the absence of $CP$ violation. In the following
sections we will only talk about the $CP$ even contributions
coming from various $\cos(j\phi)$ modulations and ignore the
discussion on $\sin(j\phi)$ modulations as they are zero.
\subsection{Spin--$\frac{1}{2}$ case: $t$--quark}
The top pair production at an $e^+e^-$ collider proceeds through a
photon and a $Z$--boson exchange in the $s$--channel. We will
study the effect of the initial polarisation of the electron
$P_{e^-}$ and positron $P_{e^+}$. The partial chiral nature of the
$Z$ coupling leads to a finite top polarisation even for
unpolarised initial state electron and positron beams. For
$t$--quark decaying into a lepton through a $W$,  the analysing
power of the top is $\alpha=1$.
\TABLE[ht]{
\renewcommand{\arraystretch}{1.3}
\begin{tabular}{||c|c|c|c|c||}\hline\hline
$(P_{e^-},P_{e^+})$ & Quantities & Frame $M$ & Frame $F$ &
Reference\\\hline \multirow{2}{*}{$(+0.00, +0.00)$} & $A^1$ &
$+0.111$ & $-0.035$ &
\multirow{2}{*}{---}\\
& $a_1/a_0$ & $+0.175$ & $-0.055$ & \\
\hline \multirow{2}{*}{$(+0.80, -0.60)$} & $A^1$ & $-0.253$ &
$+0.149$ &
\multirow{2}{*}{Fig.~\ref{fig:Tpm} }\\
& $a_1/a_0$ & $-0.397$ & $+0.234$ & \\
\hline \multirow{2}{*}{$(+0.792,+0.60)$} & $A^1$ & $\approx0$ &
$+0.021$ &
\multirow{2}{*}{Fig.~\ref{fig:Tc} }\\
& $a_1/a_0$ & $\approx0$ & $+0.033$ &\\
\hline\hline
\end{tabular}
\caption{\label{tab:top}Values for the fitted asymmetry $A^1$ and
the fit parameter $a_1/a_0$, see Eq.~\ref{eq:fita}, for the lepton
distribution from $t$-decay for different initial state
polarisations $P_{e^-},P_{e^+}$ for the electron and the positron
in frames $M$ and $F$. We have the relation $a_1/a_0 = \pi A^1/2$
which is  observed numerically within tolerance ($\pm10^{-3}$).
The other $a_j/a_0, j\neq 1$ are zero within the tolerance. Recall
that we generate $2 \; 10^6$ events.}}

We start our analysis with unpolarised beams and the polarisation
of top for this case is given as ($P_{e^\mp}$ is the polarisation
of $e^\mp$):
$$(P_{e^-},P_{e^+})=( 0.00, 0.00) : \eta_1 =+0.222, \hspace{0.5cm}
\eta_2=0.000,\hspace{0.5cm}\eta_3=-0.127 \ .$$ This corresponds to
the asymmetry $A^1=\eta_1/2=0.111$  and the amplitude of
$\cos\phi$ to be $\pi A^1/2 = 0.175$. This is confirmed by the fit
in frame M, see Table~\ref{tab:top}. We would need\footnote{
Number of events required: $N=f^2/(A^j)^2$, where $f$ is the
degree of statistical significance. Numbers with $f=3$, for
$3\sigma$ significance, are quoted.} $N_M\approx 730$ events to
measure it with $3\sigma$ significance in frame $M$. In  frame $F$
the asymmetry in this case is $A^1 = -0.035$ and requires $N_F
\approx 7350$ events for it to be measured with $3\sigma$
significance. Thus, one needs at least $\max(N_M,N_F)=7350$ events
to confirm the spin of $t$--quark to be at least $\frac{1}{2}$
with unpolarised beams. In this case where the beams are not
polarised, the asymmetries are smaller  in frame $F$ however the
analysis in this frame does  confirm that no new modulation has
been missed, and thus  reconfirms the spin--$1/2$ nature of the
top.

In order to improve the sensitivity, one might consider the case
of polarised $e^+e^-$ beams to produce top quarks with larger
polarisation. For example,
$$(P_{e^-},P_{e^+})=(+0.80, -0.60) :\hspace{0.5cm}
\eta_1 =-0.505, \hspace{0.5cm}
\eta_2=0.000,\hspace{0.5cm}\eta_3=+0.554 \ ,$$ which corresponds
to much larger polarisation and hence a larger asymmetry $A^1 =
-0.253$ in frame $M$. This requires only $N_M\approx 140$ events
to measure $A^1$ with $3\sigma$ significance. The azimuthal
distribution for this beam polarisation is shown in
Fig.~\ref{fig:Tpm} in both the frames $M$ and $F$. In frame $F$
however the asymmetry $A^1$ is smaller, see Table~\ref{tab:top},
hence we need a larger number of events, $N_F\approx410$, to
measure it with $3\sigma$ significance. Thus, one needs
$\max(N_M,N_F)=410$ events to confirm the spin of $t$--quark to be
at least $\frac{1}{2}$ with this choice of beam polarisation,
which is a large improvement over the unpolarised case. To rule
out higher asymmetries with a higher degree of significance one
still needs a larger number of events than this.
\EPSFIGURE[ht]{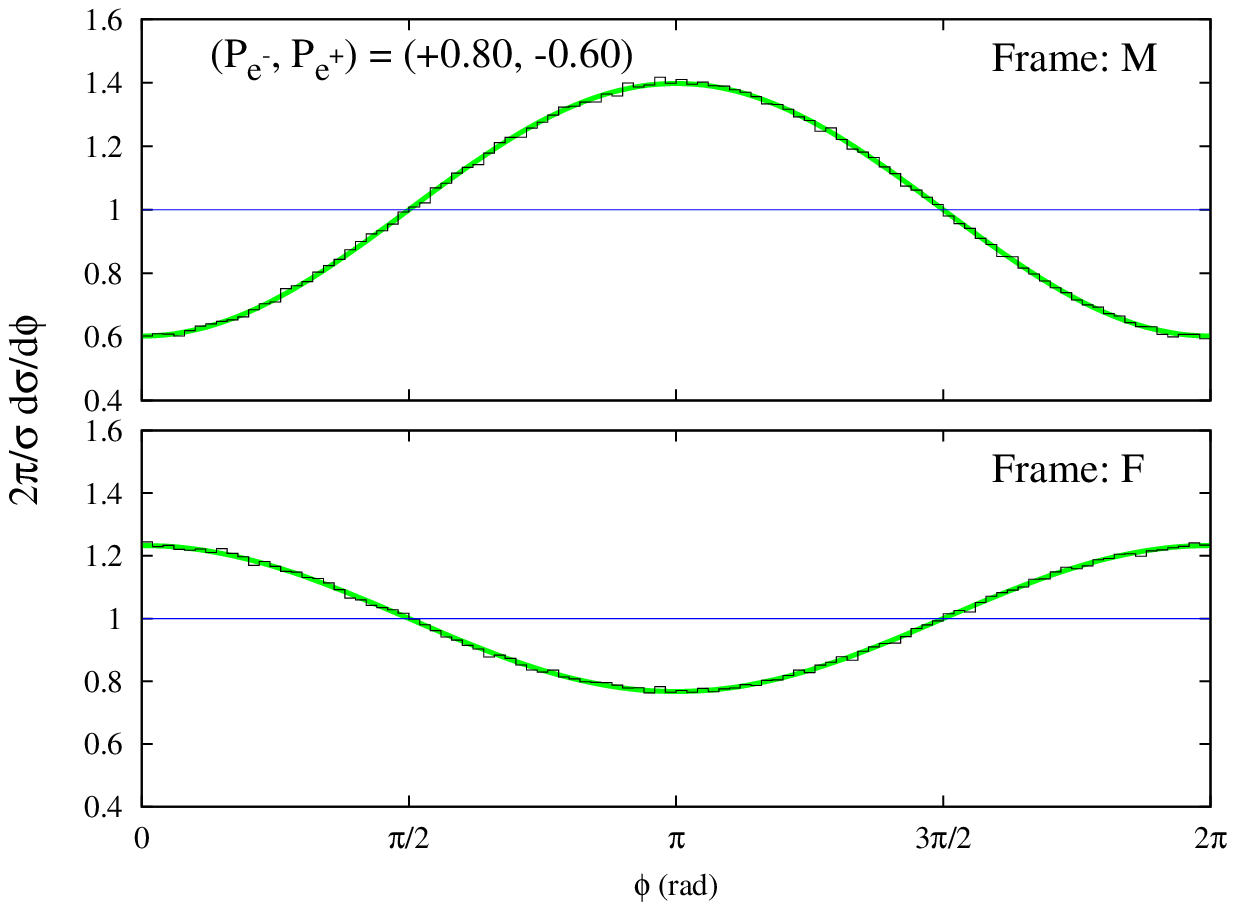} {\label{fig:Tpm}The azimuthal distribution
of lepton from decay of $t$--quark is plotted for
$(P_{e^-},P_{e^+})=(+0.80, -0.60)$ in frame $M$ (top) and in frame
$F$ (below) using $2\times10^6$ events at partonic level
(histogram). The best fit (green/grey line) to $F_4(\phi)$ leads
to the coefficient of the $\cos\phi$ modulation to be non-zero
(given in Table~\ref{tab:top}) and all other modulations are
absent in both the frames indicating the spin of $t$--quark to be
$\frac{1}{2}$. }

\EPSFIGURE[ht]{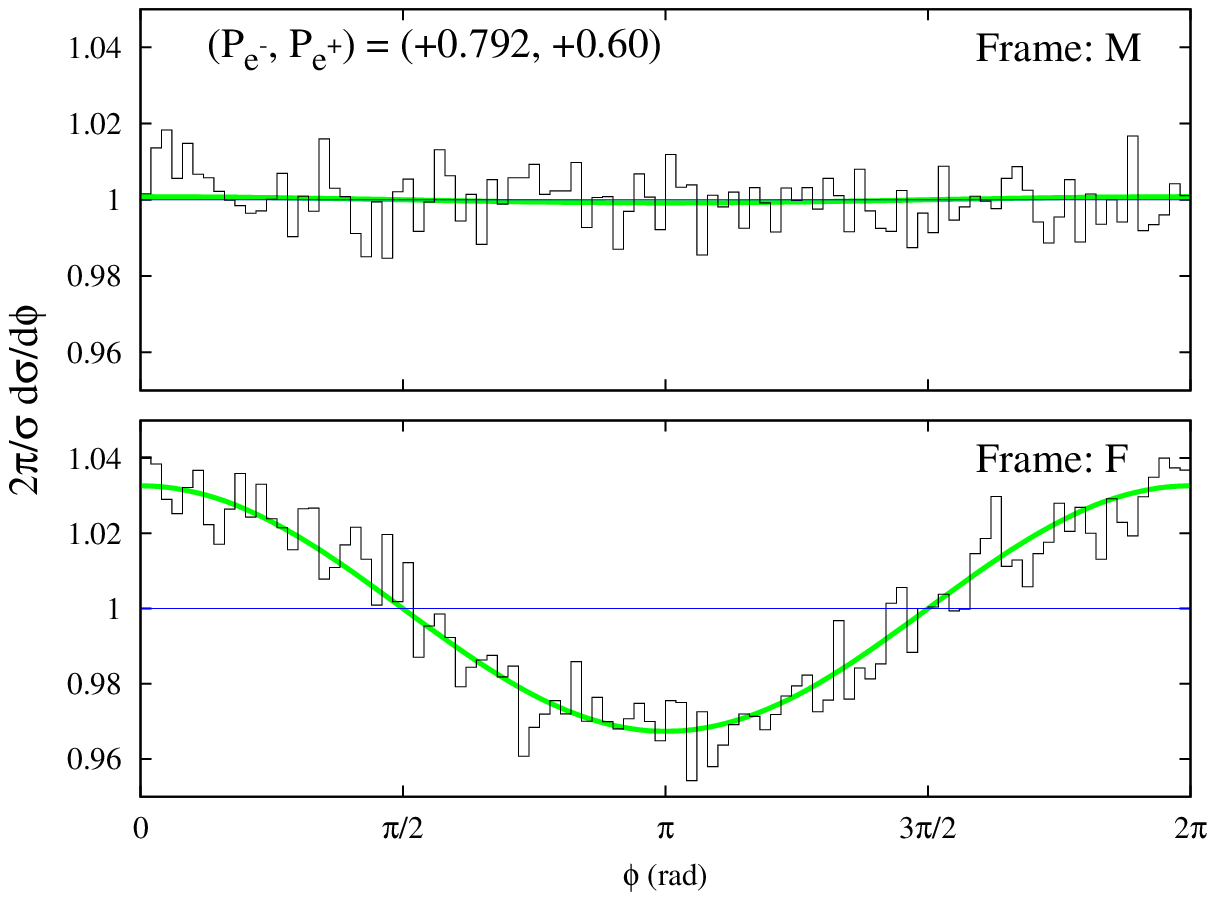} {\label{fig:Tc}The azimuthal distribution
same as in Fig.~\ref{fig:Tpm} for $(P_{e^-},P_{e^+})=(+0.792,
+0.60)$. This is special case when the distribution is flat in
frame $M$ and frame $F$ is needed for spin determination.}

Next we discuss the case when the transverse polarisation of
$t$--quark, $\eta_1$, is zero. We arrange this by tuning the beam
polarisations to appropriate values. This leads to $A^1\approx0$,
and hence in  frame $M$ the fit gives $a_1/a_0\approx 0$,  and
a flat distribution as shown in Fig.~\ref{fig:Tc}. The top
polarisations in this case are given as
$$(P_{e^-},P_{e^+})=(+0.792,+0.60) :\hspace{0.5cm}
\eta_1 =0.000, \hspace{0.5cm}
\eta_2=0.000,\hspace{0.5cm}\eta_3=+0.080 \ .$$ We note that the
longitudinal polarisation of $t$--quark, $\eta_3$, though small is
not zero and hence in frame $F$ this leads to a non-zero value of
the asymmetry $A^1$ and  the  $\cos\phi$ modulation as seen in
Fig.~\ref{fig:Tc}. In this case $N_F \approx 2 \; 10^4$ events are
required to measure this asymmetry at $3\sigma$ significance. This
example re-imposes the need for a second frame $F$ in association
with the helicity frame $M$ to measure and re-confirm the spin of
a particle.

\subsection{Spin--$1$ case: $W$--boson}
\label{sec:WMC}
\EPSFIGURE[ht]{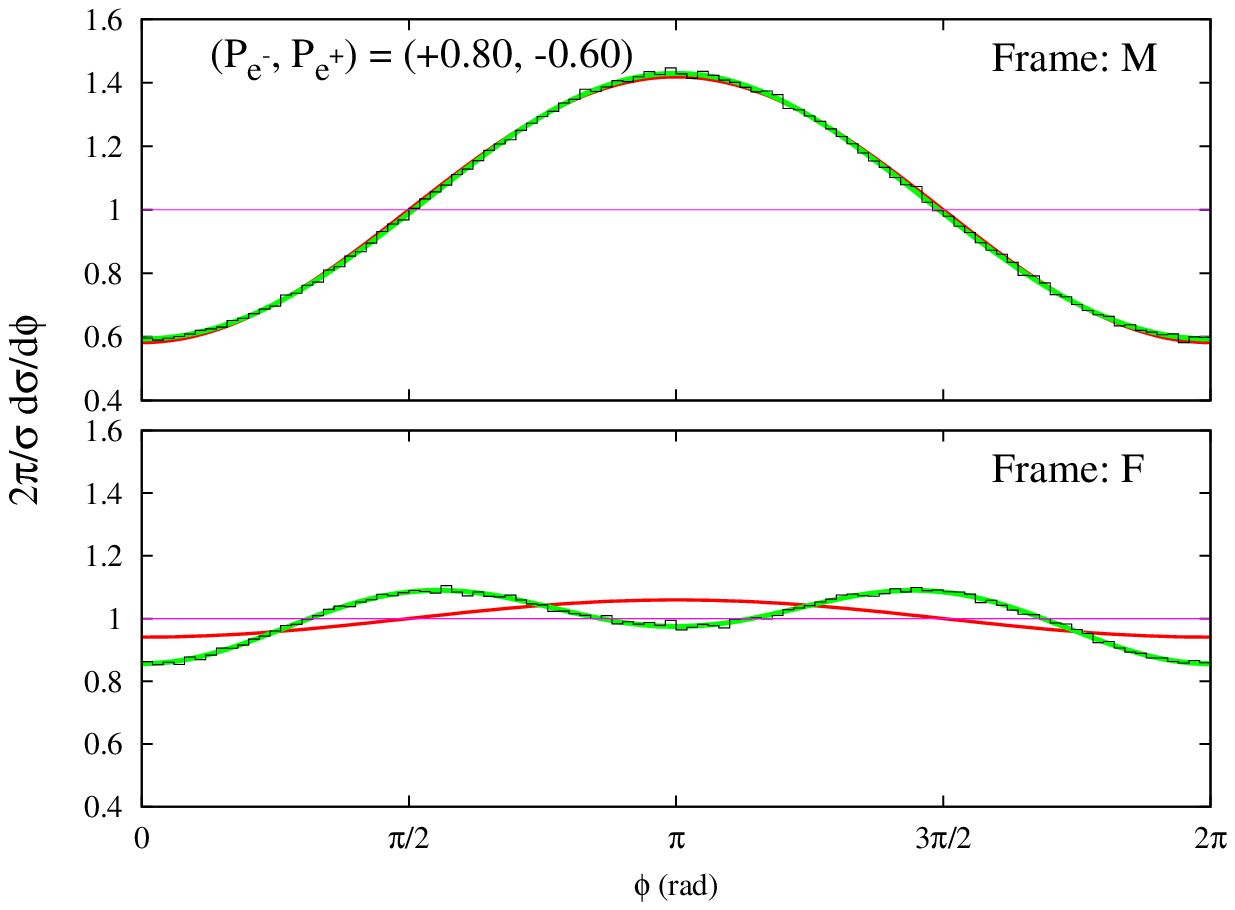} {\label{fig:Wpm}The azimuthal distribution
of lepton from decay of $W$--boson is plotted for
$(P_{e^-},P_{e^+})=(+0.80, -0.60)$ in frame $M$ (top) and in frame
$F$ (below) using $2\times10^6$ events at partonic level
(histogram). The best fit (green/grey line) to $F_4(\phi)$ leads
to the coefficient of the $\cos\phi$ and $\cos 2\phi$ modulation
to be non-zero in frame $F$ (given in Table~\ref{tab:W}) and all
other modulations are absent in both the frames indicating the
spin of $W$--boson to be $1$. The red (dark grey) line show only
the $\cos\phi$ modulation of the distribution.}
The $W$ boson analysis is a much better advocate for the need of
frame $F$, beside frame $M$. In the process under consideration,
the $W$--bosons are produced (almost) on-mass-shell as a decay
product of $t$--quark. Since the coupling of $W$--boson is chiral,
they are produced with high polarisation even in the decay of
unpolarised top quarks. For the same set of events as used for the
case of top quarks, the vector polarisations of the $W$--boson are
given by
$$(P_{e^-},P_{e^+})=(+0.80,-0.60) :\hspace{0.5cm}
p_x =+0.355, \hspace{0.5cm} p_y=0.000,\hspace{0.5cm} p_z=0.000 \
.$$ From Eq.(\ref{A2I}) we know that the coefficient of the
$\cos\phi$ modulation is proportional to $p_x$ and non-zero in
this case. The coefficient of $\cos2\phi$ modulation is
proportional to tensor polarisation $(T_{xx}- T_{yy})$, which
happens to be zero\footnote{We note that the asymmetry $A^2$ is
zero in the helicity frame $M$ for on-shell $W$ bosons, but
numerically we find it to be small but non-zero as the decay width
of $W$ is not very small and there is a non-negligible
contribution from off-shell $W$s.} for this process in the
helicity frame $M$.
\TABLE[ht]{
\renewcommand{\arraystretch}{1.3}
\begin{tabular}{||c|c|c|c|c||}\hline\hline
$(P_{e^-},P_{e^+})$ & Quantities & Frame $M$ & Frame $F$ &
Reference\\\hline \multirow{4}{*}{$(+0.80, -0.60)$} & $A^1$ &
$-0.266$ & $-0.038$ &
\multirow{4}{*}{Fig.~\ref{fig:Wpm} }\\
& $A^2$ & $\approx0$ & $-0.054$ & \\
& $a_1/a_0$ & $-0.418$ & $-0.059$ & \\
& $a_2/a_0$ & $\approx0$ & $-0.086$ & \\
\hline \multirow{4}{*}{$(+0.75,+0.60)$} & $A^1$ & $\approx0$ &
$-0.093$ &
\multirow{4}{*}{Fig.~\ref{fig:WW} }\\
& $A^2$ & $\approx0$ & $-0.026$ & \\
& $a_1/a_0$ & $\approx0$ & $-0.147$ &\\
& $a_2/a_0$ & $\approx0$ & $-0.041$ &\\
\hline\hline
\end{tabular}
\caption{\label{tab:W}The table of asymmetries $A^1 \ \& \ A^2$
and the fit parameter $a_1/a_0 \ \& \ a_2/a_0$ for lepton's
distribution from $W$--boson decay  for different initial state
polarisations in frames $M$ and $F$. We have the relation $a_i/a_0
= \pi A^i/2$, which is also observed numerically within tolerance
($\pm10^{-3}$). }}
This leads to only $\cos\phi$ modulation of the azimuthal
distribution as seen in Fig.~\ref{fig:Wpm} for frame $M$. Using
the helicity amplitudes given in Appendix~\ref{ap:ffv}, one can
write the production density matrix for $W$--boson, which is
produced in the decay of $t$--quark, and we easily see that
$\rho_W(+1,-1)=\rho_W(-1,+1)=0$ in the helicity frame $M$ due to
angular momentum conservation. Here, a higher spin (spin--$1$)
particle is produced in the decay process of lower spin
(spin--$\frac{1}{2}$), thus it can~not span all its helicity
states for fixed helicities of other particles and hence leads to
$\rho_W(\pm1,\mp1)=0$. This is proven for the general case in
Appendix~\ref{ap:high} in the helicity frame. However, in the
boosted frame $F$ the asymmetry $A^2$ measures
$T^F_{xx}-T^F_{yy}$, which is non-zero in general, see
Eq.(\ref{eq:pT_F}). In frame $F$ we find $A^2=-0.054$ which leads
to a $\cos2\phi$ modulation of the azimuthal angle in this frame,
see Fig.~\ref{fig:Wpm}. Here $N_F=3100$ events will be required to
measure $A^2$ with $3\sigma$ significance. Further, all the higher
$A^j$s ($A^{j>2}$) are found to be zero in both  frames proving
that the particle under consideration to be spin--$1$ and its
production process to be $CP$-conserving. The asymmetries and fit
parameters are listed in Table~\ref{tab:W} for both the frames.

Next we look at a case where the azimuthal distribution in the
helicity frame $M$ is flat which would wrongly suggest that the
particle is a scalar. The various vector polarisations are given
as
$$(P_{e^-},P_{e^+})=(+0.75,+0.60) :\hspace{0.5cm}
p_x =+0.000, \hspace{0.5cm} p_y=0.000,\hspace{0.5cm} p_z=0.193 \
.$$ In the helicity frame $M$, the asymmetry $A^2$ is zero due to
the angular momentum conservation and $A^1$ is zero because it is
proportional to $p_x$, which is zero for the chosen initial state
beam polarisations. The $W$--boson appears to be spin--$0$ in this
frame $M$ with this particular beam polarisations. The asymmetries
$A^j$ and the fit parameters $a_j/a_0$ are listed in
Table~\ref{tab:W} for this case and the corresponding azimuthal
distributions are plotted in Fig.~\ref{fig:WW}. Changing over to
frame $F$ leads to non-zero values of both $A^1$ and $A^2$, see
Table~\ref{tab:W}, and the corresponding azimuthal distribution
visibly has the $\cos2\phi$ modulation, Fig.~\ref{fig:WW}. In this
case $N_F\approx1.3\times 10^4$ events are required to measure
$A^2$ with $3\sigma$ significance. This is the best example of a
case where one needs a frame other than the helicity frame to
confirm the spin of the particle, which is polarised with
$p_z\neq0$ and $T_{zz}\neq 0$. We, however, note that this process
is not the best process to study the spin of $W$-boson. For this
purpose one should look at the pair production process $e^+e^-\to
W^+W^-$ as discussed in Ref.~\cite{Buckley:2008pp}.
\EPSFIGURE[ht]{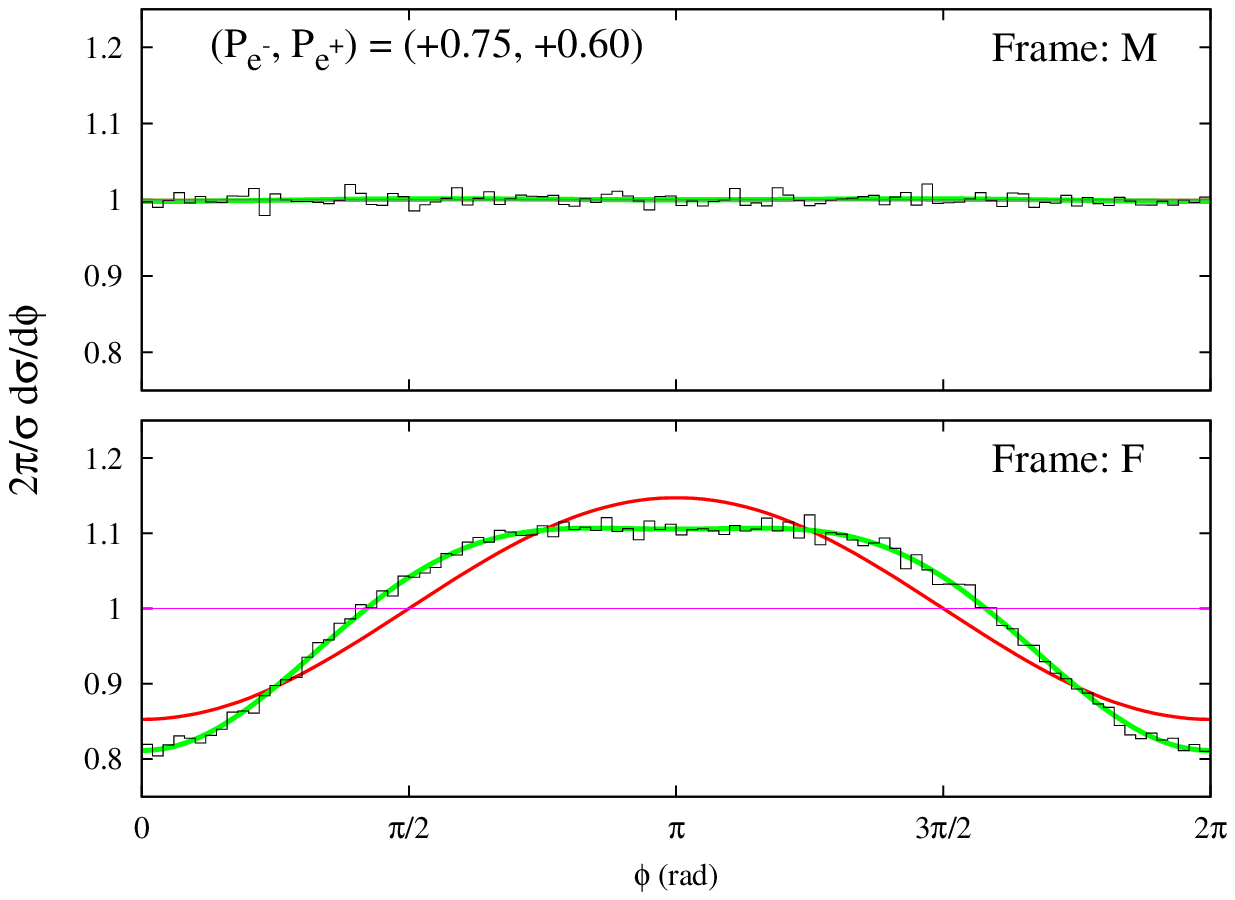} {\label{fig:WW}The azimuthal distribution
same as in Fig.~\ref{fig:Wpm} for $(P_{e^-},P_{e^+})=(+0.75,
+0.60)$. This is special case where frame $M$ has flat
distribution and frame $F$ is needed for the spin determination. }

Thus we conclude that one needs two different reference frames to
measure and re-confirm the spin of a particle using the same set
of events. Further, if the event set includes a cascade decay, one
can construct the asymmetries for different particles using the
spin vector $s_i$ for different particles and using the momentum
$p_B$ of different final state particles. For example, in the
above case, we could have used the momentum of the $b$--quark to
construct the asymmetries in place of the leptons. In the events
with hadronic decay of $W$s, one could use either of the jets to
construct the correlators and hence the asymmetries. Thus, using
different final state particles, we can find a larger set of
asymmetries to re-confirm the spin of a particle, however we can
not improve the significance of the measurement by combining
different correlators for the same set of events. A larger event
sample is necessary to improve the statistical significance of the
measurements.

\section{Discussions and Conclusions}
In this paper, we constructed observables to measure the spin of a
heavy unstable particle produced at a collider by harvesting the
spin dependence of the azimuthal distribution through the quantum
interference between  the different helicity states, {\it i.e} the
non diagonal elements of the helicity density matrix. The aim is
to construct observables that are sensitive to the highest
rank--$2s$ tensor polarisation of a particle with spin--$s$ that
lead to  a $\cos 2 s \phi$ modulation of the azimuthal
distribution. Such a method has been known for a long time and we
have provided an analytical understanding of it. In particular,
the novelty of our approach is the construction of two reference
frames where in one of them the spin basis is subjected to a Wick
rotation. The latter mixes the longitudinal polarisation and the
transverse polarisation in the production plane and therefore the
spin modulation in the azimuthal angle is sensitive to this
mixture whereas in the standard approach the longitudinal
polarisation is integrated away and does not contribute to the
usual azimuthal asymmetries. The construction of two frames allows
within the same experiment and with the same event sample to cross
check the spin measurement based on azimuthal asymmetries. In some
cases this can be crucial since  the usual transverse polarisation
tensor/vector can be zero either accidentally or for a dynamical
reason and therefore would lead to a wrong conclusion. This can be
rescued in the second frame provided the longitudinal polarisation
is not zero as well. We have shown examples in the decay of the
top and the $W$ where this occurs with SM production and decay
mechanisms. In the appendices we consider more general couplings
and decays than those that describe the SM particles. This helps
in obtaining a set of conditions on the production and decay
mechanisms for some of the asymmetries to be  non-zero and hence
the spin to be measured. One drawback of the method however, as
outlined in the present paper, is that it requires complete
reconstruction of the test particle's momentum which is necessary
to build up the needed spin vectors. If there are too many
invisible particles this might not be possible especially in a
machine like the LHC where the partonic centre of mass energy is
not fixed. The same drawback also affects other methods of spin
reconstruction. We feel however that is  worth investigating how
the method we have described can be exploited in combination with
other methods or by making some mild assumption on the spectrum of
the event or the underlying physics.

\acknowledgments We thank G.~B\'elanger, R.~Godbole, M. Guichait
and S.~Rindani for the many fruitful discussions during the
initial phase of the project as well as W.~Porod later on. We also
thank R.~Godbole, W.~Porod and S.~Rindani for reading the
manuscript and making useful comments. This work was partially
supported by CEFIPRA under IF-3004-B and  the French ANR project,
{\tt ToolsDMColl}. The work of RKS was partially supported by the
Initiative and Networking Fund of the Helmholtz Association,
contract HA-101 {\tt Physics at the Terascale}.

\appendix
%
\section{Rotation matrices $d^j_{m,n}(\theta)$}
The general form of the $d$ function is given in Eq.(\ref{eq:ds}).
For completeness and although these can be easily found in many
textbooks, we explicitly write the $d$ function up to spin--$2$.
To avoid clutter we take as short-hand notation $c= \cos
(\theta/2)$ and $s= \sin (\theta/2)$ then all the
$d^j_{m,n}$ useful to our study are given below.\\[0.2cm]
$\bullet$ $j=0$ : $d^0_{0,0}$ = $1$\\[0.2cm]
$\bullet$ $j=\frac{1}{2}$ : $d^{\frac{1}{2}}_{m,n}$ = $\left[
\renewcommand{\arraystretch}{1.2}
\begin{tabular}{cc}
$c$ & $-s$\\
$s$ & $ c$
\end{tabular} \right] $ \\[0.2cm]
$\bullet$ $j=1$ : $d^{1}_{m,n}$ = $\left[
\renewcommand{\arraystretch}{1.2}
\begin{tabular}{ccc}
$c^2$ & $-\sqrt{2}cs$ & $s^2$ \\
$\sqrt{2}cs$ & $2c^2-1$ & $-\sqrt{2}cs$ \\
$s^2$ & $\sqrt{2}cs$ & $c^2$
\end{tabular} \right] $ \\[0.2cm]
$\bullet$ $j=\frac{3}{2}$ : $d^{\frac{3}{2}}_{m,n}$ =  $\left[
\renewcommand{\arraystretch}{1.2}
\begin{tabular}{cccc}
$c^3$ & $-\sqrt{3}sc^2$ & $\sqrt{3}s^2c$ & $-s^3$\\
$\sqrt{3}sc^2$ & $c-3s^2c$ & $s-3sc^2$ & $\sqrt{3}s^2c$\\
$\sqrt{3}s^2c$ & $-s+3sc^2$ & $c-3s^2c$ & $-\sqrt{3}sc^2$\\
$s^3$ & $\sqrt{3}s^2c$ & $\sqrt{3}sc^2$ & $c^3$
\end{tabular} \right] $ \\[0.2cm]
$\bullet$ $j=2$ : $d^{2}_{m,n}$ = \\$\left[
\renewcommand{\arraystretch}{1.2}
\begin{tabular}{ccccc}
$c^4$ & $-2sc^3$ & $\sqrt{6}s^2c^2$ & $-2s^3c$ & $s^4$\\
$2sc^3$ &$-3c^2+4c^4$ & $\sqrt{6}sc(s^2-c^2)$ & $3s^2-4s^4$ & $-2s^3c$\\
$\sqrt{6}s^2c^2$ & $-\sqrt{6}sc(s^2-c^2)$ & $1-6c^2+6c^4$ &
$\sqrt{6}sc(s^2-c^2)$ & $\sqrt{6}s^2c^2$\\
$2s^3c$ & $3s^2-4s^4$ & $-\sqrt{6}sc(s^2-c^2)$ & $-3c^2+4c^4$ & $-2sc^3$\\
$s^4$ & $2s^3c$ & $\sqrt{6}s^2c^2$ & $2sc^3$ & $c^4$
\end{tabular} \right] $ \\

\section{Decay density matrix for higher spin particle}
\label{ap:dden} As a short-hand notation we now define $C=
\cos\theta$ and $S= \sin\theta$ which enter the expressions for
the density matrices of higher  spin particles, namely $s=3/2$ and
$s=2$ briefly discussed in the main text. The corresponding
normalised decay matrices are calculated from
Eq.~\ref{eq:norm-ddm} and using the explicit expressions for the
$d$ matrices.
\subsection{Spin-${\frac{3}{2}}$ particle}
For the decay $|\frac{3}{2},l\rangle \to |s_1,l_1\rangle +
|s_2,l_2\rangle$, the decay density matrix is given by
\begin{eqnarray}
\Gamma^{\frac{3}{2}}(+\frac{3}{2},+\frac{3}{2})&=&
\frac{(1+2\gamma_1)+3(\alpha_1+\alpha_2) C +3(1-2\gamma_1) C ^2
+(\alpha_2-3\alpha_1) C ^3}{8} \nonumber\\
\Gamma^{\frac{3}{2}}(+\frac{3}{2},+\frac{1}{2})&=& \frac{\sqrt{3}
\  S  \ [ \ (\alpha_1+\alpha_2)+2(1-2\gamma_1)
 C  + (\alpha_2-3\alpha_1) C ^2 \ ]}{8} \ e^{i\phi} \nonumber\\
\Gamma^{\frac{3}{2}}(+\frac{3}{2},-\frac{1}{2})&=& \frac{\sqrt{3}
\  S ^2 \ [ \ (1-2\gamma_1)+(\alpha_2-3\alpha_1)
 C  \ ]}{8} \ e^{i 2\phi}\nonumber\\
\Gamma^{\frac{3}{2}}(+\frac{3}{2},-\frac{3}{2})&=&
\frac{(\alpha_2-3\alpha_1) S ^3}{8} \ e^{i 3\phi} \nonumber\\
\Gamma^{\frac{3}{2}}(+\frac{1}{2},+\frac{3}{2})&=& \frac{\sqrt{3}
\  S  \ [ \ (\alpha_1+\alpha_2)+2(1-2\gamma_1)
 C  + (\alpha_2-3\alpha_1) C ^2 \ ]}{8} \ e^{-i\phi} \nonumber\\
\Gamma^{\frac{3}{2}}(+\frac{1}{2},+\frac{1}{2})&=&
\frac{(3-2\gamma_1)+(3\alpha_2-5\alpha_1)C - 3(1-2\gamma_1)C^2
-3(\alpha_2-3\alpha_1) C^3  \ ]}{8} \nonumber\\
\Gamma^{\frac{3}{2}}(+\frac{1}{2},-\frac{1}{2})&=& \frac{S \ [ \
(3\alpha_2-\alpha_1) -3(\alpha_2-3\alpha_1)C^2 \ ]}{8}
 \ e^{i\phi}\nonumber\\
\Gamma^{\frac{3}{2}}(+\frac{1}{2},-\frac{3}{2})&=& \frac{\sqrt{3}
\  S ^2 \ [ \ (1-2\gamma_1) -(\alpha_2-3\alpha_1)
 C  \ ]}{8} \ e^{i2\phi} \nonumber\\
\Gamma^{\frac{3}{2}}(-\frac{1}{2},+\frac{3}{2})&=& \frac{\sqrt{3}
\  S ^2 \ [ \ (1-2\gamma_1)+(\alpha_2-3\alpha_1)
 C  \ ]}{8} \ e^{-i2\phi}\nonumber\\
\Gamma^{\frac{3}{2}}(-\frac{1}{2},+\frac{1}{2})&=& \frac{S \ [ \
(3\alpha_2-\alpha_1) -3(\alpha_2-3\alpha_1)C^2 \ ]}{8}
 \ e^{-i\phi}\nonumber\\
\Gamma^{\frac{3}{2}}(-\frac{1}{2},-\frac{1}{2})&=&
\frac{(3-2\gamma_1)-(3\alpha_2-5\alpha_1)C - 3(1-2\gamma_1)C^2
+3(\alpha_2-3\alpha_1) C^3  \ ]}{8} \nonumber\\
\Gamma^{\frac{3}{2}}(-\frac{1}{2},-\frac{3}{2})&=& \frac{\sqrt{3}
\  S  \ [ \ (\alpha_1+\alpha_2)-2(1-2\gamma_1)
 C  + (\alpha_2-3\alpha_1) C ^2 \ ]}{8} \ e^{i\phi} \nonumber\\
\Gamma^{\frac{3}{2}}(-\frac{3}{2},+\frac{3}{2})&=&
\frac{(\alpha_2-3\alpha_1) S ^3}{8} \ e^{-i3\phi} \nonumber\\
\Gamma^{\frac{3}{2}}(-\frac{3}{2},+\frac{1}{2})&=& \frac{\sqrt{3}
\  S ^2 \ [ \ (1-2\gamma_1) -(\alpha_2-3\alpha_1)
 C  \ ]}{8} \ e^{-i2\phi} \nonumber\\
\Gamma^{\frac{3}{2}}(-\frac{3}{2},-\frac{1}{2})&=& \frac{\sqrt{3}
\  S  \ [ \ (\alpha_1+\alpha_2)-2(1-2\gamma_1)
 C  + (\alpha_2-3\alpha_1) C ^2 \ ]}{8} \ e^{-i\phi}
\nonumber\\
\Gamma^{\frac{3}{2}}(-\frac{3}{2},-\frac{3}{2})&=&
\frac{(1+2\gamma_1) -3(\alpha_1+\alpha_2) C +3(1-2\gamma_1) C ^2
-(\alpha_2-3\alpha_1) C ^3}{8}
\nonumber\\
\end{eqnarray}
where,
\begin{eqnarray}
\alpha_1=\frac{a^{3/2}_{1/2}-a^{3/2}_{-1/2}}{\sum_l
a^{3/2}_l},\hspace{0.6cm}
\alpha_2=\frac{a^{3/2}_{3/2}-a^{3/2}_{-3/2}}{\sum_l
a^{3/2}_l},\hspace{0.6cm}
\gamma_1=\frac{a^{3/2}_{1/2}+a^{3/2}_{-1/2}}{\sum_l a^{3/2}_l}
\end{eqnarray}
and
\begin{eqnarray}
a^{3/2}_{3/2}&=&\frac{1}{\pi}\sum_{l_1} |{\cal
M}^{3/2}_{l_1,l_1-\frac{3}{2}}|^2  \hspace{1.0cm}
|l_1| \le s_1, \ \ |l_1-\frac{3}{2}| \le s_2 \nonumber\\
a^{3/2}_{1/2}&=&\frac{1}{\pi}\sum_{l_1} |{\cal
M}^{3/2}_{l_1,l_1-\frac{1}{2}}|^2  \hspace{1.0cm}
|l_1| \le s_1, \ \ |l_1-\frac{1}{2}| \le s_2 \nonumber\\
a^{3/2}_{-1/2}&=&\frac{1}{\pi}\sum_{l_1} |{\cal
M}^{3/2}_{l_1,l_1+\frac{1}{2}}|^2  \hspace{1.0cm}
|l_1| \le s_1, \ \ |l_1+\frac{1}{2}| \le s_2 \nonumber\\
a^{3/2}_{-3/2}&=&\frac{1}{\pi}\sum_{l_1} |{\cal
M}^{3/2}_{l_1,l_1+\frac{3}{2}}|^2  \hspace{1.0cm}
|l_1| \le s_1, \ \ |l_1+\frac{3}{2}| \le s_2 \nonumber\\
\end{eqnarray}
\subsection{Spin-$2$ particle}
For the decay $|2,l\rangle \to |s_1,l_1\rangle + |s_2,l_2\rangle$,
the decay density matrix is given by
\begin{eqnarray}
\Gamma^2(+2,+2)&=& \left[ \ A_0 + 4A_1 C  + 6A_2 C ^2 + 4A_3  C ^3
+ A_4  C ^4
 \ \right] \nonumber\\
\Gamma^2(+2,+1)&=& 2\left[ \ A_1 + 3A_2 C  + 3 A_3  C ^2 +  A_4 C
^3
 \ \right] \  S  \ e^{i\phi} \nonumber\\
\Gamma^2(+2,+0)&=& \sqrt{6}\left[ \ A_2 + 2A_3  C  + A_4 C ^2
 \ \right] \  S ^2 \ e^{i2\phi} \nonumber\\
\Gamma^2(+2,-1)&=& 2\left[ \ A_3 +  A_4  C
 \ \right] \  S ^3 \ e^{i3\phi} \nonumber\\
\Gamma^2(+2,-2)&=&
 A_4
 \  S ^4 \ e^{i4\phi} \nonumber\\
\Gamma^2(+1,+2)&=& 2\left[ \ A_1 + 3A_2 C  + 3 A_3  C ^2 +  A_4 C
^3
 \ \right] \  S  \ e^{-i\phi} \nonumber\\
\Gamma^2(+1,+1)&=& 4\left[ \ 1 + 2(A_1-3\beta)  C  - 3 A_2  C ^2 -
2A_3 C ^3
-A_4 C ^4 \ \right] \nonumber\\
\Gamma^2(+1,+0)&=& 2\sqrt{6}\left[ \ 2(\beta + A_2 C ) +   S ^2 \
(A_3 + A_4 C )
\ \right] \  S  \ e^{i\phi} \nonumber\\
\Gamma^2(+1,-1)&=& 4\left[ \ 3A_2 + A_4 S ^2
\ \right] \  S ^2 \ e^{i2\phi} \nonumber\\
\Gamma^2(+1,-2)&=& 2\left[ \ A_3 - A_4 C
\ \right] \  S ^3 \ e^{i3\phi} \nonumber\\
\Gamma^2(+0,+2)&=& \sqrt{6}\left[ \ A_2 + 2A_3  C  + A_4 C ^2
 \ \right] \  S ^2 \ e^{-i2\phi} \nonumber\\
\Gamma^2(+0,+1)&=& 2\sqrt{6}\left[ \ 2(\beta + A_2 C ) +   S ^2 \
(A_3 + A_4 C )
\ \right] \  S  \ e^{-i\phi} \nonumber\\
\Gamma^2(+0,+0)&=& 4\left[ \ 4\delta + 3 A_2  S ^2 + 3 A_4  S ^4
 \ \right] \nonumber\\
\Gamma^2(+0,-1)&=& 2\sqrt{6}\left[ \ 2(\beta - A_2 C ) +   S ^2 \
(A_3 - A_4 C )
\ \right] \  S  \ e^{i\phi} \nonumber\\
\Gamma^2(+0,-2)&=& \sqrt{6}\left[ \ A_2 - 2A_3  C  + A_4 C ^2
 \ \right] \  S ^2 \ e^{i2\phi} \nonumber\\
\Gamma^2(-1,+2)&=& 2\left[ \ A_3 +  A_4  C
 \ \right] \  S ^3 \ e^{-i3\phi} \nonumber\\
\Gamma^2(-1,+1)&=& 4\left[ \ 3A_2 + A_4 S ^2
\ \right] \  S ^2 \ e^{-i2\phi} \nonumber\\
\Gamma^2(-1,+0)&=& 2\sqrt{6}\left[ \ 2(\beta - A_2 C ) +   S ^2 \
(A_3 - A_4 C )
\ \right] \  S  \ e^{-i\phi} \nonumber\\
\Gamma^2(-1,-1)&=& 4\left[ \ 1 - 2(A_1-3\beta)  C  - 3 A_2  C ^2 +
2A_3 C ^3
-A_4 C ^4 \ \right] \nonumber\\
\Gamma^2(-1,-2)&=& 2\left[ \ A_1 - 3A_2 C  + 3 A_3  C ^2 -  A_4 C
^3
 \ \right] \  S  \ e^{i\phi} \nonumber\\
\Gamma^2(-2,+2)&=&
 A_4
 \  S ^4 \ e^{-i4\phi} \nonumber\\
\Gamma^2(-2,+1)&=& 2\left[ \ A_3 - A_4 C
\ \right] \  S ^3 \ e^{-i3\phi} \nonumber\\
\Gamma^2(-2,+0)&=& \sqrt{6}\left[ \ A_2 - 2A_3  C  + A_4 C ^2
 \ \right] \  S ^2 \ e^{-i2\phi} \nonumber\\
\Gamma^2(-2,-1)&=& 2\left[ \ A_1 - 3A_2 C  + 3 A_3  C ^2 -  A_4 C
^3
 \ \right] \  S  \ e^{-i\phi} \nonumber\\
\Gamma^2(-2,-2)&=& \left[ \ A_0 - 4A_1 C  + 6A_2 C ^2 - 4A_3  C ^3
+ A_4  C ^4
 \ \right]
\end{eqnarray}
where,
\begin{eqnarray}
A_0 &=& \frac{a^2_2+4a^2_1+6a^2_0+4a^2_{-1}+a^2_{-2}}
{16 \ \sum_la^2_l}, \nonumber \\
A_1 &=& \frac{a^2_2+2a^2_1-2a^2_{-1}-a^2_{-2}}
{16 \ \sum_la^2_l}, \nonumber \\
A_2 &=& \frac{a^2_2-2a^2_0+a^2_{-2}}
{16 \ \sum_la^2_l}, \nonumber \\
A_3 &=& \frac{a^2_2-2a^2_1+2a^2_{-1}-a^2_{-2}}
{16 \ \sum_la^2_l}, \nonumber \\
A_4 &=& \frac{a^2_2-4a^2_1+6a^2_0-4a^2_{-1}+a^2_{-2}}
{16 \ \sum_la^2_l}, \nonumber \\
\beta &=& \frac{ a^2_1-a^2_{-1}} {\sum_la^2_l}, \hspace{0.5cm}
\delta= \frac{ a^2_0} {\sum_la^2_l} \label{eq:A4}
\end{eqnarray}
and
\begin{eqnarray}
a^{2}_{2}&=&\frac{5}{4\pi}\sum_{l_1} |{\cal M}^{2}_{l_1,l_1-2}|^2
\hspace{1.0cm}
|l_1| \le s_1, \ \ |l_1-2| \le s_2 \nonumber\\
a^{2}_{1}&=&\frac{5}{4\pi}\sum_{l_1} |{\cal M}^{2}_{l_1,l_1-1}|^2
\hspace{1.0cm}
|l_1| \le s_1, \ \ |l_1-1| \le s_2 \nonumber\\
a^{2}_{0}&=&\frac{5}{4\pi}\sum_{l_1} |{\cal M}^{2}_{l_1,l_1}|^2
\hspace{1.0cm}
|l_1| \le \min{s_1, s_2} \nonumber\\
a^{2}_{-1}&=&\frac{5}{4\pi}\sum_{l_1} |{\cal M}^{2}_{l_1,l_1+1}|^2
\hspace{1.0cm}
|l_1| \le s_1, \ \ |l_1+1| \le s_2 \nonumber\\
a^{2}_{-2}&=&\frac{5}{4\pi}\sum_{l_1} |{\cal M}^{2}_{l_1,l_1+2}|^2
\hspace{1.0cm} |l_1| \le s_1, \ \ |l_1+2| \le s_2.
\end{eqnarray}
%
%
\section{Helicity amplitudes and the analysing power}
\label{ap:hel} In this section we give expressions for the
helicity amplitudes  pertaining to  2--body decay processes of
spin--$\frac{1}{2}$ and spin--$1$ particles. The expressions are
derived for a general dimension-4 effective operator describing
the  coupling of the particles. This will permit to give the
different analysing power coefficients.

We take the mass of the mother particle to be $m$ and that of
daughters to be $m_1$ and $m_2$, the polar and azimuthal angle
belongs to the first particle with mass $m_1$. The energy and the
momentum of the daughter particles are given as
\begin{eqnarray}
E_1=\frac{m^2+m_1^2-m_2^2}{2m},\hspace{0,5cm}
E_2=\frac{m^2+m_2^2-m_1^2}{2m},\nonumber\\
p=\frac{\sqrt{((m+m_2)^2-m_1^2)((m+m_1)^2-m_2^2)}}{2m}
\label{eq:2bd}
\end{eqnarray}
from 2--body decay kinematics. Below we discuss the 2--body decay
of a fermion and a vector boson into two massive particles.
\subsection{Decay: $|\frac{1}{2},\lambda\rangle \to |\frac{1}{2},\lambda_1
\rangle + |1,\lambda_2\rangle$} \label{ap:ffv} For this decay the
helicity for the fermion $\lambda=\pm 1/2$ will be denoted as
$\lambda=\pm 1/2$ and for the bosons $\lambda=\pm 1$ as
$\lambda=\pm $ such that the helicity
$M(\lambda,\lambda_1,\lambda_2)$ writes
as  $M(+,+,+) = M(+\frac{1}{2},+\frac{1}{2},+1)$.\\
\noindent The decay vertex it taken to be $\bar f_1 \gamma^\mu \ (C_L P_L
+ C_R P_R) f_2 V_\mu$ with real $C_{L,R}$ and the amplitudes are
listed below in the rest frame of the decaying particle:
\begin{eqnarray}
M(+,+,+)&=& \left[-(C_L \ P_1^-     -C_R \ P_1^+     )\right] \
e^{\frac{+i\phi}{2}} \  \left(-\sin\frac{\theta}{2}\right)  \nonumber\\
M(+,+,0)&=& \left[-(C_L \ P_1^-P_2^--C_R \ P_1^+P_2^+)\right] \
e^{\frac{+i\phi}{2}} \ \left(+\cos\frac{\theta}{2}\right) \nonumber\\
M(+,+,-)&=&0\nonumber\\
M(+,-,+)&=&0\nonumber\\
M(+,-,0)&=& \left[+(C_L \ P_1^+P_2^+-C_R \ P_1^-P_2^-)\right] \
e^{\frac{+i\phi}{2}} \ \left(-\sin\frac{\theta}{2}\right) \nonumber\\
M(+,-,-)&=& \left[+(C_L \ P_1^+     -C_R \ P_1^-     )\right] \
e^{\frac{+i\phi}{2}} \ \left(+\cos\frac{\theta}{2}\right) \nonumber\\
M(-,+,+)&=& \left[-(C_L \ P_1^-     -C_R \ P_1^+     )\right] \
e^{\frac{-i\phi}{2}} \ \left(+\cos\frac{\theta}{2}\right) \nonumber\\
M(-,+,0)&=& \left[-(C_L \ P_1^-P_2^--C_R \ P_1^+P_2^+)\right] \
e^{\frac{-i\phi}{2}} \ \left(+\sin\frac{\theta}{2}\right) \nonumber\\
M(-,+,-)&=&0\nonumber\\
M(-,-,+)&=&0\nonumber\\
M(-,-,0)&=& \left[+(C_L \ P_1^+P_2^+-C_R \ P_1^-P_2^-)\right] \
e^{\frac{-i\phi}{2}} \ \left(+\cos\frac{\theta}{2}\right) \nonumber\\
M(-,-,-)&=& \left[+(C_L \ P_1^+     -C_R \ P_1^-     )\right] \
e^{\frac{-i\phi}{2}} \ \left(+\sin\frac{\theta}{2}\right) \nonumber\\
\end{eqnarray}
Here, the terms in the square brackets are the reduced matrix
elements, $({\cal M}^s_{\lambda_1,\lambda_2}/\sqrt{2\pi})$ and the
$d^s_{\lambda, \lambda_1-\lambda_2}$ functions are enclosed in
round brackets. The symbols $P_{1,2}^\pm$ are defined as
\begin{equation}
P_1^\pm = \sqrt{m} \ \ \frac{E_1+m_1\pm p}{\sqrt{E_1+m_1}},
\hspace{0.5cm} P_2^\pm = \frac{1}{\sqrt{2}} \sqrt{\frac{E_2\pm
p}{E_2\mp p}}. \label{eq:P12}
\end{equation}
Using the expressions of the reduced matrix elements, the
analysing power $\alpha$ for this decay can be written as
\begin{equation}
\alpha=\frac{(C_R^2-C_L^2)(1-x_1^2-2x_2^2)\sqrt{1+(x_1^2-x_2^2)^2-2(x_1^2+x_2^2)
}}{(C_R^2+C_L^2)(1-2x_1^2+x_2^2+x_1^2x_2^2+x_1^4-2x_2^4)-12C_LC_Rx_1x_2^2}
\ \ , \label{eq:alffv}
\end{equation}
where $x_i=m_i/m$. For the decay of top quark, $t\to bW$, with
$m_1=m_b=0$ within the SM we have $C_R=0$ leading to
$\alpha=-(1-2x_2^2)/(1+2x_2^2) \sim -0.38$.
\subsection{Decay: $|\frac{1}{2},\lambda\rangle \to |\frac{1}{2},\lambda_1
\rangle + |0,0\rangle$} \label{ap:ffs} As done in the previous
section the helicity for the fermions $\lambda=\pm 1/2$ will be
denoted as $\lambda=\pm 1/2$ such that the helicity
$M(\lambda,\lambda_1$ writes as $M(+,+)$ for
$M(+\frac{1}{2},+\frac{1}{2})$.

For this decay the helicity amplitudes, $M(\lambda,\lambda_1)
=M(+,+) = M(+\frac{1}{2},+\frac{1}{2})$. The decay vertex it taken
to be $\bar f_1 \gamma^\mu \ (C_L P_L + C_R P_R)f_2 \ S$ with
complex $C_{L,R}$ and all the amplitudes are listed below:
\begin{eqnarray}
M(+,+) &=&\left[\frac{C_R \ P_1^- +C_L \ P_1^+}{\sqrt{2}}\right]
 \ e^{\frac{+i\phi}{2}} \ \left(+\cos\frac{\theta}{2}\right) \nonumber \\
M(+,-) &=&\left[\frac{C_R \ P_1^+ +C_L \ P_1^-}{\sqrt{2}}\right]
 \ e^{\frac{+i\phi}{2}} \ \left(-\sin\frac{\theta}{2}\right) \nonumber \\
M(-,+) &=&\left[\frac{C_R \ P_1^- +C_L \ P_1^+}{\sqrt{2}}\right]
 \ e^{\frac{-i\phi}{2}} \ \left(+\sin\frac{\theta}{2}\right) \nonumber \\
M(-,-) &=&\left[\frac{C_R \ P_1^+ +C_L \ P_1^-}{\sqrt{2}}\right]
 \ e^{\frac{-i\phi}{2}} \ \left(+\cos\frac{\theta}{2}\right)
\end{eqnarray}
Here, the reduced matrix elements, $({\cal
M}^s_{\lambda_1,0}/\sqrt{2\pi})$, are given in square brackets and
the  $d^s_{\lambda,\lambda_1}$ functions in the round brackets.
The symbols $P_1^\pm$ are same as in Eq.(\ref{eq:P12}). Using the
expressions of reduced matrix elements we get the expression for
$\alpha$, the analysing power of the spin--$\frac{1}{2}$ particle,
as
\begin{equation}
\alpha =
\frac{-(|C_R|^2-|C_L|^2)\sqrt{1+(x_1^2-x_2^2)^2-2(x_1^2+x_2^2)}}
{(|C_R|^2+|C_L|^2)(1+x_1^2-x_2^2) + 4x_1 \Re(C_LC_R^*)} \ \ ,
\label{eq:alffs}
\end{equation}
where $x_i=m_i/m$. Thus we need the scalar to have parity
violating couplings, $|C_L|\neq|C_R|$, for the analysing power to
be non-zero. However, for a neutral scalar, say the neutral Higgs
boson of the MSSM, we have $C_L=C_R^*$ in other words $|C_L|
=|C_R|$ leading to $\alpha=0$. The same occurs in a CP conserving
MSSM with any of the neutral Higgs boson. Thus we should chose
processes involving squarks for spin measurement in the decay of
gauginos.
\subsection{Decay: $|1,\lambda\rangle \to |\frac{1}{2},\lambda_1
\rangle + |\frac{1}{2},\lambda_2\rangle$} \label{ap:vff} We take
the same convention as in~\ref{ap:ffv} with the same operator for
the interaction. We find
\begin{eqnarray}
M(+,+,+) &=&\left[-i \frac{C_R \ p_1^+p_2^- +C_L \
p_1^-p_2^+}{2}\right]
 \ e^{+i\phi} \ \left(\frac{-1}{\sqrt{2}} \sin\theta\right) \nonumber \\
M(+,+,-) &=&\left[+i \frac{C_R \ p_1^+p_2^+ +C_L \
p_1^-p_2^-}{\sqrt{2}}\right]
 \ e^{+i\phi} \ \left( \cos^2\frac{\theta}{2}\right) \nonumber \\
M(+,-,+) &=&\left[-i \frac{C_R \ p_1^-p_2^- +C_L \
p_1^+p_2^+}{\sqrt{2}}\right]
 \ e^{+i\phi} \ \left( \sin^2\frac{\theta}{2}\right) \nonumber \\
M(+,-,-) &=&\left[+i \frac{C_R \ p_1^-p_2^+ +C_L \
p_1^+p_2^-}{2}\right]
 \ e^{+i\phi} \ \left(\frac{-1}{\sqrt{2}} \sin\theta\right) \nonumber \\
M(0,+,+) &=&\left[-i \frac{C_R \ p_1^+p_2^- +C_L \
p_1^-p_2^+}{2}\right]
 \ \left(\cos\theta\right) \nonumber \\
M(0,+,-) &=&\left[+i \frac{C_R \ p_1^+p_2^+ +C_L \
p_1^-p_2^-}{\sqrt{2}}\right]
 \ \left( \frac{+1}{\sqrt{2}} \sin\theta\right) \nonumber \\
M(0,-,+) &=&\left[-i \frac{C_R \ p_1^-p_2^- +C_L \
p_1^+p_2^+}{\sqrt{2}}\right]
 \ \left( \frac{-1}{\sqrt{2}} \sin\theta\right) \nonumber \\
M(0,-,-) &=&\left[+i \frac{C_R \ p_1^-p_2^+ +C_L \
p_1^+p_2^-}{2}\right]
 \ \left(\cos\theta\right) \nonumber \\
M(-,+,+) &=&\left[-i \frac{C_R \ p_1^+p_2^- +C_L \
p_1^-p_2^+}{2}\right]
 \ e^{-i\phi} \ \left(\frac{+1}{\sqrt{2}} \sin\theta\right) \nonumber \\
M(-,+,-) &=&\left[+i \frac{C_R \ p_1^+p_2^+ +C_L \
p_1^-p_2^-}{\sqrt{2}}\right]
 \ e^{-i\phi} \ \left( \sin^2\frac{\theta}{2}\right) \nonumber \\
M(-,-,+) &=&\left[-i \frac{C_R \ p_1^-p_2^- +C_L \
p_1^+p_2^+}{\sqrt{2}}\right]
 \ e^{-i\phi} \ \left( \cos^2\frac{\theta}{2}\right) \nonumber \\
M(-,-,-) &=&\left[+i \frac{C_R \ p_1^-p_2^+ +C_L \
p_1^+p_2^-}{2}\right]
 \ e^{-i\phi} \ \left(\frac{+1}{\sqrt{2}} \sin\theta\right)
 \label{eq:appdec1toh}
\end{eqnarray}
In Eq.~\ref{eq:appdec1toh} the terms in the square bracket stand
for the reduced matrix elements $({\cal
M}^s_{\lambda_1,\lambda_2}\sqrt{\frac{3}{4\pi}})$, the terms in
the round  brackets are the $d^s_{\lambda,\lambda_1-\lambda_2}$
functions. The symbols $p_{1,2}^\pm$  are defined as
\begin{equation}
p_1^\pm = \frac{E_1+m_1\pm p}{\sqrt{E_1+m_1}},\hspace{0.5cm}
p_2^\pm = \frac{E_2+m_2\pm p}{\sqrt{E_2+m_2}},
\end{equation}
Using the expressions for $a^s_l$ and reduced matrix element we
get expressions for two parameter $\alpha$ and $\delta$ as
\begin{eqnarray}
\alpha&=&\frac{2(C_R^2-C_L^2)\sqrt{1+(x_1^2-x_2^2)^2-2(x_1^2+x_2^2)}}
{12 C_LC_R x_1x_2+(C_R^2+C_L^2)[2-(x_1^2-x_2^2)^2+(x_1^2+x_2^2)]},\\
\delta&=&\frac{4C_LC_R x_1x_2
+(C_R^2+C_L^2)[(x_1^2+x_2^2)-(x_1^2-x_2^2)^2]} {12 C_LC_R
x_1x_2+(C_R^2+C_L^2)[2-(x_1^2-x_2^2)^2+(x_1^2+x_2^2)]},
\end{eqnarray}
where $x_i=m_i/m$. \\
\noindent If the final state fermions are massless, $x_1\to0,x_2\to0$, one obtains
$\alpha\to(C_R^2-c_L^2)/(C_R^2+C_L^2)$ and $\delta\to0$. This is
the case for the decay of $W$ and $Z$ bosons into massless
fermions. Further, for the decay of $W$s, within the SM we have
$C_R=0$ hence $\alpha=-1$.
\subsection{Decay: $|1,\lambda\rangle \to |1,\lambda_1\rangle + |0,0\rangle$}
\label{ap:vvs} For this decay the helicity amplitudes,
$M(\lambda,\lambda_1)=M(+,+)=M(+1,+1)$. The decay vertex it taken
to be $C_{VVS} g_{\mu\nu}V^\mu V_1^\nu $ with real $C_{VVS}$ and
the helicity amplitudes are given by:
\begin{eqnarray}
M(+,+)&=&\left[-C_{VVS} \right] \ e^{+i\phi} \
\left(\cos^2\frac{\theta}{2}\right) \nonumber\\
M(+,0)&=&\left[-C_{VVS} \frac{E_1}{m_1}\right] \ e^{+i\phi} \
\left(\frac{-\sin\theta}{\sqrt{2}}\right)  \nonumber\\
M(+,-)&=&\left[-C_{VVS} \right] \ e^{+i\phi} \
\left(\sin^2\frac{\theta}{2}\right) \nonumber\\
M(0,+)&=& \left[-C_{VVS} \right] \ \left(\frac{\sin\theta}{\sqrt{2}} \right)
\nonumber\\
M(0,0)&=&\left[-C_{VVS} \frac{E_1}{m_1}\right] \ \left(\cos\theta \right)
 \nonumber\\
M(0,-)&=&\left[-C_{VVS} \right] \ \left(\frac{-\sin\theta}{\sqrt{2}} \right)
\nonumber\\
M(-,+)&=&\left[-C_{VVS} \right] \ e^{-i\phi} \
\left(\sin^2\frac{\theta}{2}\right)  \nonumber\\
M(-,0)&=&\left[-C_{VVS} \frac{E_1}{m_1}\right] \ e^{-i\phi} \
\left(\frac{\sin\theta}{\sqrt{2}}\right) \nonumber\\
M(-,-)&=&\left[-C_{VVS} \right] \ e^{-i\phi} \
\left(\cos^2\frac{\theta}{2}\right)
\end{eqnarray}
This leads to $a^1_1=a^1_{-1}$ hence $\alpha=0$. $\delta$ is given
by
\begin{equation}
\delta=\frac{(1+x_1^2-x_2^2)^2}{1+(x_1^2-x_2^2)^2+2(5x_1^2-x_2^2)}.
\end{equation}
Such decays occur  the models of extra-dimensions, for example,
$W^{(1)} \to W^{(0)} h$, $Z^{(1)} \to Z^{(0)} h$ etc.
\subsection{Decay: $|1,\lambda\rangle \to |1,\lambda_1\rangle + |1,
\lambda_2\rangle$} \label{ap:vvv} For this decay the helicity
amplitudes, $M(\lambda,\lambda_1,\lambda_2)
=M(+,+,+)=M(+1,+1,+1)$. The decay vertex is taken to be $C_{VVV}
T_{\mu\nu\rho} V^\mu V_1^\nu V_2^\rho$ where $C_{VVV}$ is real. We
only assume here a standard {\em gauge} tri-linear coupling
$$T_{\mu\nu\rho}=\left[ g_{\mu\nu}(q-p_1)_\rho + g_{\nu\rho}(p_1-p_2)_\mu +
g_{\rho\mu}(p_2-q)_\nu \right]$$ with $q$ being the 4-momentum of
the mother particle and $p_{1,2}$ is the 4-momentum of the
daughter particles. All momenta are assumed incoming at the
interaction vertex. With this notation the non-zero helicity
amplitudes are given by:
\begin{eqnarray}
M(+,+,+)&=&\left[2 C_{VVV} \; p\right] \ e^{+i\phi} \
\left(\frac{-\sin\theta}{\sqrt{2}}\right) \nonumber\\
M(+,0,0)&=&\left[-2 C_{VVV} \; p \ \frac{E_1E_2+p^2}{m_1m_2} \right] \
e^{+i\phi} \ \left(\frac{-\sin\theta}{\sqrt{2}}\right) \nonumber\\
M(+,-,-)&=&\left[2 C_{VVV} \; p\right] \ e^{+i\phi} \
\left(\frac{-\sin\theta}{\sqrt{2}}\right) \nonumber\\
M(0,+,+)&=& \left[2 C_{VVV} \; p\right] \ \left(\cos\theta\right) \nonumber\\
M(0,0,0)&=& \left[-2 C_{VVV} \; p \ \frac{E_1E_2+p^2}{m_1m_2} \right] \
\left(\cos\theta\right) \nonumber\\
M(0,-,-)&=& \left[2 C_{VVV} \; p\right] \  \left(\cos\theta\right) \nonumber\\
M(-,+,+)&=&\left[2 C_{VVV} \; p\right] \ e^{-i\phi} \
\left(\frac{-\sin\theta}{\sqrt{2}}\right) \nonumber\\
M(-,0,0)&=&\left[-2 C_{VVV} \; p \ \frac{E_1E_2+p^2}{m_1m_2} \right] \
e^{-i\phi} \ \left(\frac{-\sin\theta}{\sqrt{2}}\right) \nonumber\\
M(-,-,-)&=&\left[2 C_{VVV} \; p\right] \ e^{-i\phi} \
\left(\frac{-\sin\theta}{\sqrt{2}}\right)
\end{eqnarray}
This leads to $a^1_1=a^1_{-1}=0$ and hence $\alpha=0$ and
$\delta=1$. Example of such decays, in the models of
extra-dimensions, are $W^{(1)} \to W^{(0)} Z^{(0)}$, $W^{(1)} \to
W^{(0)} \gamma^{(0)}$ etc.

Above we saw that the parameters $\alpha$ and $\delta$ have a
simple expressions in terms of masses and couplings of the
particles involved. This can be simply added to a spectrum
generation code, such as {\tt SOFTSUSY}~\cite{Allanach:2001kg},
{\tt SuSpect}~\cite{Djouadi:2002ze}, {\tt
SPheno}~\cite{Porod:2003um} for SUSY models, and one can quickly
know which decay channel is the best for estimation of the
particle's spin.
%
%
\section{Higher spin particle disguising as lower spin particle }
\label{ap:high} If a higher spin particle is produced as a decay
product of the lower spin particle, its spin orientations are
restricted. This makes the  particle appear as of a lower spin in
frame $M$ see Sec.~\ref{sec:WMC}. Since the total differential
rate is the product of production and decay density matrices,
$$ d\sigma = \frac{1}{2I} \ \rho^s(l,l') \ \times \ \Gamma^s(l,l') \ \
d\Phi_n \ .$$ Thus, for the decay distribution to have a $2s\phi$
modulation, we must have $\rho^s(s,-s)=\rho^{s*}(-s,s) \neq 0$.
Now if this spin $s$ particle were produced in the decay reaction
$|j,m\rangle \to |s_1,l_1\rangle + |s,l\rangle$, then the
production density matrix is given by
\begin{eqnarray}
\rho^s(l,l') &=& \sum_{m,l_1} \ M^{jm}_{l_1,l} \  M^{jm*}_{l_1,l'}\nonumber\\
&=& \left(\frac{2j+1}{4\pi}\right) e^{i(l-l')\phi} \sum_{m,l_1} \
d^j_{m,l_1-l} \ d^j_{m,l_1-l'} \ \ {\cal M}^j_{l_1,l} {\cal
M}^{j*}_{l_1,l'}.
\end{eqnarray}
Thus we have extreme off-diagonal term given by
\begin{eqnarray}
\rho^s(s,-s) &=& \left(\frac{2j+1}{4\pi}\right) e^{i2s\phi}
\sum_{m,l_1} \ d^j_{m,l_1-s} \ d^j_{m,l_1+s} \ \ {\cal
M}^j_{l_1,l} {\cal M}^{j*}_{l_1,l'}.
\end{eqnarray}
For this to be non-zero, we must have
$$ |l_1-s| \le j \ \ {\rm and} \ \ |l_1+s|\le j $$
for at least one value of $l_1$. However, this condition is never
satisfied (for any $l_1$) when we have $s>j$, i.e.
$\rho^s(s,-s)=0$ for $s>j$. This leads to the absence of the
highest mode in the  $\phi$ distribution in frame $M$. This is
numerically demonstrated for the sample of $W$ boson production
from the decay of $t$-quark. Since the helicities are invariant
only under the boost along the momentum (which does not changes
the direction of the momentum), the density matrix goes through a
similarity transformation when boosted in any other direction.
Thus in frame $F$, in general one can have a non-zero value for
$\rho^s(s,-s)$ and hence the $2s\phi$ modulation of the azimuthal
distribution.
%
%


\end{document}